\documentclass[aps,showpacs,showkeys,onecolumn]{revtex4}
\usepackage{amsmath,amssymb} 
\usepackage{graphicx} 
\begin{document}
\title{The non-perturbative  renormalization group in the ordered phase.}
\date{\today}
\author{\firstname{Jean-Michel} \surname{Caillol}}
\email{Jean-Michel.Caillol@th.u-psud.fr}
\affiliation{Univ. Paris-Sud, Laboratoire LPT, UMR8627, Orsay, F-91405, France}
\affiliation{CNRS, Orsay, F-91405, France}
\begin{abstract}
We study some analytical properties of the  solutions of  the  non perturbative renormalization group 
flow  equations
for  a  scalar field theory with $Z_2$ symmetry in the ordered phase, \textit{i.e.} at temperatures below the 
critical temperature.
The study is made in the framework of the local potential approximation.
We show that the  required physical discontinuity of 
the magnetic susceptibility $\chi(M)$ at $M=\pm M_0$ ($M_0$ spontaneous magnetization) is reproduced
only if the cut-off function which separates high and low energy modes 
satisfies  to some restrictive explicit mathematical conditions; we stress that  these conditions are not satisfied 
by a sharp cut-off in dimensions of space $d<4$.

By generalizing
a method proposed  earlier by Bonanno and Lacagnina  
( \textit{Nucl. Phys. B} \textbf{693}  (2004) 36.) to any kind of
cut-off  we propose to solve numerically the   renormalization group  flow equations for 
the threshold functions rather than for  the local potential.
It yields
an  algorithm sufficiently robust and precise to extract  universal as well as non universal quantities
from numerical experiments at any temperature, in particular at sub-critical temperatures  in the ordered phase.
Numerical  results obtained for the $\varphi^4$ potential with three different cut-off functions are
 reported and compared.
The data  confirm our theoretical predictions concerning the analytical behavior of  $\chi(M)$ at $M=\pm M_0$. 

Fixed point solutions of the adimensionned  renormalization group 
flow equations are also obtained  in the same vein, that is by solving
the fixed points equations and the associated eigenvalue problem for the threshold functions rather than for the potential.
We report high precision data for  the odd and even spectra of critical exponents  for  different cut-offs obtained in this way.

\end{abstract} 
\maketitle
\section{\label{sec:intro}Introduction}
During the last twenty years the non-perturbative approach  to the renormalization group (RG)
originated by Wilson~\cite{Wilson,Wegner}
has been the subject of a revival in both  statistical and quantum field theory.
Two main formulations of the non perturbative renormalization group (NPRG) have been developed in parallel
to study a system at equilibrium at, or near to,  criticality ; for instance, in the simplest case, a system described, 
at microscopic scale $\Lambda$ (in momentum
scale),  by an action  $\mathcal{S}_{\Lambda}[\varphi]$ where $\varphi$ is a scalar field. 
In the first approach, one  realizes a continuous  RG transformation of the action $\mathcal{S}_k [\varphi]$ 
 from $k=\Lambda$ to $k=0$ and, \textit{a priori},  no expansion  with respect to whatsoever small parameter
being required. 
At scale-$k$ (in momentum space) the high energy
modes $\widetilde{\varphi}_q$, $\arrowvert q \arrowvert>k$, are integrated out in the ``Wilsonian'' action
$\mathcal{S}_k[\varphi]$  which is a functional of the slow modes  $\widetilde{\varphi}_q$, $\arrowvert q\arrowvert <k$.  
This ''coarse-graining'' operation
requires the implementation of some cut-off  of the propagator, either sharp or soft, aiming at separating slow ($\arrowvert q\arrowvert <k$) 
and fast ($\arrowvert q\arrowvert >k$) modes.  This ''coarse-graining'' process is devised in such a way  that all  
the Wilsonian actions $\mathcal{S}_k[\varphi]$ yield
the same physics in the infra-red (IR) limit ($q \to 0 $). 
The flow of  $\mathcal{S}_k[\varphi]$,  from the microscopic  scale
$k=\Lambda$ to the macroscopic scale $k=0$, is governed by  the
Wilson-Polchinski equation~\cite{Wilson,Wegner,Polchinski}
in case of a smooth cut-off and the  Wegner-Houghton~\cite{WH,Hasen} equation in case of a sharp cut-off. 
Powerfull approximation schemes has been devised to obtain approximate, non-perturbative solutions of
these equations; for a review see Ref.~\cite{Bervillier}. 

The second  formulation, called the ``effective average action'' approach, was developed after the seminal works 
of Nicoll,  Chang and Stanley for 
the sharp cut-off version~\cite{Ni1,Ni2} and Wetterich, Ellwanger and Morris for the smooth cut-off 
version~\cite{Wetterich00,Wetterich0,Ellwanger,Morris}. This method
implements on the effective average action $\Gamma_k[\phi]$ - roughly speaking the Gibbs free energy of the fast modes 
$\widetilde{\varphi}_q$, $q>k$ of the
classical field   $\phi=\langle \varphi\rangle$  - rather than on  the Wilsonian action $\mathcal{S}_k[\varphi]$ 
the ideas of integration of high-energy modes that underlies any RG approach. 
The flow of $\Gamma_k$
results in  equations which can be solved under the same kind of non perturbative approximations than 
those used for the Wilson-Polchinski or Wegner-Houghton equations. 
The main advantage of this more recent formulation is that it gives access to  the RG flow of physical quantities, 
i.e. the Gibbs free energy $\Gamma_k[\phi]$ and its field derivatives, the vertex functions, rather than to such  an  abstract 
object  as the Wilsonian action $\mathcal{S}_k[\varphi]$. 
Quite remarkably, the same kind of formalism was developed in a less elaborated language by Reatto \textit{et al.} in the domain of the theory of classical 
liquids \cite{ReattoI, ReattoII,Reatto0}, many years before these recent contributions to statistical field theory.
 The relations between  these  corpora of works is discussed in ref.~\cite{Caillol_Li}. 
Recent reviews and  lectures devoted to Wetterich's approach are
available~\cite{Wetterich,Delamotte} and should be consulted for a thorough discussion.
Wetterich's version of the RG  is  in fact equivalent to that of Wilson-Polchinski  as  discussed  in 
Ref.~\cite{Morris,Caillol_RG}. 

In this paper we will adhere  to Wetterich point of view and focus on the study of approximate solutions of the NPRG 
for   a scalar field theory  with $Z_2$ symmetry,   at a temperature $T$ below the critical temperature 
$T_{c}$, i.e. in the ``ordered phase'' or
``two phase region''.
For $T< T_c$ the system exhibits a spontaneous magnetization $M = <\varphi>$
which can take any value in the interval $(- M_0(T), M_0(T))$ with an equal probability in the thermodynamic limit.
Henceforth we shall reserve the name spontaneous magnetization for $ M_0(T)$.
As a result the susceptibility $\chi$ is infinite, or equivalently,  the second derivative of the Gibbs free energy with respect to
the magnetization 

$\Gamma^{''}(M)=\chi^{-1}=0$ in the ordered phase. The potential $\Gamma (M)$, strictly convex for $\arrowvert  M \arrowvert > M_0$
($\Gamma^{''}(M)=\chi^{-1} > 0$) is affine for $- M_0 \leq M \leq M_0$.

The simplest non-perturbative version of the NPRG, \textit{i.e.} the local potential approximation (LPA),  yields indeed
a convex free energy with a plateau as noticed in Refs~\cite{ReattoI,ReattoII,Reatto0,Wetterich,Bonanno}. 
However it has been discovered
by Reatto \textit{et al.} that the  analytic behavior of $\Gamma^{''}(M)$ in the vicinity of $\pm M_0$ depends strongly on the
choice of cut-off and dimension $d$.
For a sharp cut-off (and in $d=3$), $\Gamma^{''}(M)$ is a continuous function of  $M$, notably  at $\pm M_0$, and thus $\Gamma^{''}(M_0 +) =0$
\cite{ReattoI}.
This is a serious flaw of the theory since obviously $\Gamma^{''}(M)$  should be discontinuous at  $\pm M_0$,  \textit{i.e.} vanishing identically
in the two phase region with a jump to a finite positive value
outside the two-phase region corresponding to a finite susceptibility. In other words one should have $\Gamma^{''}(M_0 -) = 0$ and 
$\Gamma^{''}(M_0 +) > 0$. However,  with another choice of cut-off function proposed
by Litim \cite{Litim}, it was shown  by Parolla \textit{et al.} in  Ref.~\cite{ReattoII}  that, at least in $d=3$,  $\Gamma^{''}(M)$ exhibits the correct discontinuity at 
$M= \pm M_0$.
 In this work we extend these results to any kind of cut-off and 
dimensions $d$ of space and obtain the mathematical properties to be satisfied by the cut-off function to obtain the required discontinuity
of the susceptibility at $M = \pm M_0$. The crux of the whole matter is that the RG flow of Gibbs free energy  stops in the ordered phase at some finite RG time and
that the solutions of the RG flow can thus  be obtained  as asymptotic stationnary solutions for well chosen dimensionned functions
the behavior of which gives insight on the region $M \sim M_0$.

These analytical results are then checked by numerical experiments in the case of a $\varphi^4$ potential
where the RG flow equations are solved with a new algorithm which generalizes
 that devised by Bonanno and Lacagnina~\cite{Bonanno} to any kind of cut-off.
The idea is to solve the RG flow for the threshold function rather 
than for the potential itself, or one of its derivatives. The  partial differential equation for the threshold function
is of quasi-linear parabolic
type (rather than of non-linear parabolic type for the potential yielding huge numerical instabilities in the ordered phase) and can thus be solved with an arbitrary high accuracy.
The approach to convexity of the Gibbs free energy is then achieved exactly
with $\Gamma^{''}(M)$ vanishing identically in the ordered phase, down  to the smallest real available on your computer if wanted. 
Solving RG flow equations for the threshold function above $T_c$ is also possible of course, but of less interest.
A good numerical  precision can therefore 
be achieved making possible to extract precise universal as well as non-universal quantities from numerical experiments.
Thorough numerical investigations have thus been
undertaken with three different cut-off functions,  \textit{i.e.} the sharp cut-off, Litim's regulator and an 
exponential smooth cut-off,  aiming \textit{inter alia} at
testing the theoretical predictions on the behavior of $\chi(M)$ at $M=\pm M_0$; a thorough comparison and discussion
of the data provided by the different cut-off is also presented.

This study is finally completed by solving the adimensionned flow
equations asymptotically  in the same vein as that used to solve the dimensionned equations, \text{i.e.} by solving the fixed point equations
and the associated eigenvalue problem for the threshold functions rather than for the potential.
Fixed point functions and critical
exponents are then obtained for the  three cut-offs with a very high numerical precision, notably for the exponential smooth cut-off non considered 
up to now; in the case of the sharp and Litim's cut-off we recover the results obtained previously \cite{BerGiacoI,BerGiacoII}.

The paper is organized as follows; in section~\ref{Flow} we summarize briefly Wetterich formalism and the 
LPA approximation. The properties
of the RG in the two-phase region are then explored in depth yielding the mathematical conditions to be fulfilled by the cut-off function in order
to obtain the correct discontinuous behavior of $\Gamma^{''}(M)$ at $M=\pm M_0$. In section~\ref{numerI} we devise a new numerical algorithm
in which a correct account of both initial and boundary conditions is given. This material is used to perform extensive numerical experiments aimed
at testing the theoretical predictions of section~\ref{Flow} and at computing some universal and non-universal quantities of the $\varphi^4$ potential.
Section~\ref{adim} is devoted to a somehow new presentation and numerical study of the adimensionned flow equations. Estimations
of the critical exponents of the $\varphi^4$ (or Ising) $3d$ model in the LPA and for three different cut-off are reported.   In the case of Litim's and 
sharp cut-off our results are in perfect agreement with
those of ref.\cite{BerGiacoI,BerGiacoII} and  in the case of  the exponential smooth cut-off we provide the first high precision data for the even and odd spectrum of critical exponents. We conclude
 in  Section~\ref{Conclu}

\section{\label{Flow}NPRG Flow equations in the local potential approximation}
\subsection{\label{MOD} The model and the NPRG}
We consider  a scalar field theory defined at scale $\Lambda$  by it's microscopic action 
\begin{equation}
 S_{\Lambda}\left[ \varphi\right]  =
 \int_{x} \, \left\{  \left(  \vec{\nabla} \varphi_{x} \right) ^{2} + V_{\Lambda}
\left(\varphi_{x} \right)  \right\} \; ,
\end{equation} 
where $\varphi_{x}$ is a scalar field defined at some point $x$ of the domain $\Omega$ which
encloses the system, $ \int_{x}\equiv \int_{\Omega} d^{d}x$ ($d \equiv $ dimensions of space.),   $Z_2$ 
symmetry is assumed (i.e. $ V_{\Lambda}\left(\varphi_x \right)=V_{\Lambda}\left(-\varphi_x \right)$)
and  we stick to the convention that brackets $\left[ \ldots \right] $  denote functionals while
parenthesis $\left( \ldots \right)$ are for functions. Below, for numerical applications, we shall
restrict ourselves to   the $\varphi^4$ potential , i.e.   a coarse-grained version of the Ising model with
$V_{\Lambda}\left(\varphi_{x} \right)=\frac{1}{2!} r \varphi_{x} ^{2} +
 \frac{1}{4!}g \varphi_{x} ^{4}$ ( $g >0$), and  will denote by $r_c$ the value of parameter $r$ at the critical point
in the LPA approximation. 
Finally it is understood that $\Lambda$ acts as an ultra-violet (UV)  cut-off in integrals defined
in Fourier space, i.e. $ \int_{q} \equiv  \int_{ \arrowvert q \arrowvert < \Lambda} d^{d}q/(2 \pi)^d$.

Following Wetterich~\cite{Wetterich} we introduce a family of related $k$-models depending on an index $k$ 
in momentum space with $0\leq k \leq \Lambda$.  The bare action $ S_{k}\left[ \varphi\right] $
of the $k-$model - not to be confused with the Wilsonian action briefly alluded to in the introduction- is defined by
\begin{equation}
 \label{Sk}
S_{k}\left[ \varphi\right] = S_{\Lambda}\left[ \varphi\right]  +
\frac{1}{2} \varphi \cdot \mathcal{R}_{k} \cdot \varphi  \; ,
\end{equation} 
where  $\varphi \cdot \mathcal{R}_{k} \cdot \varphi \equiv \int_{p, q} \widetilde{\varphi}_{p} \,
 \widetilde{ \mathcal{R}}_{k}(p,q)  \, \widetilde\varphi_{q} $  is a massive term
aimed at separating high and low energy field modes. $ \widetilde{ \mathcal{R}}_{k}(p,q)$ is called
the cut-off or regulator function. Translational invariance implies
$\widetilde{\mathcal{R}}_{k}(p,q)=\widetilde{R}_{k}(q^2)(2\pi)^{d} \delta^{d}(p+q) $ and  the general
behavior of the cut-off function is  $\widetilde{R}_{k}(q^2) =Z_k k^2 (1 - \Theta_{\epsilon}(q, k))$ where
$Z_k$ is some positive prefactor accounting for field-renormalization and
$\Theta_{\epsilon}(q, k))$ is some smooth approximation of the Heaviside  step
 function $\Theta(q-k)$ so that when  $\epsilon \to 0$ then
 $\Theta_{\epsilon}(q, k) \to \Theta(q-k)$. The cut-off functions must have  the following generic behavior 
\begin{itemize}
 \item[(i)] when $k=0$, $\widetilde{R}_{k=0}(q^2) =0$ identically ($\forall q$) and the original model is recovered.
\item[(ii)] when $k=\Lambda$, $\widetilde{R}_{k=\Lambda}(q^2) = Z_{\Lambda}\Lambda^2 $ is sufficiently ''large '' so that  all the fluctuations
are frozen by the massive term, i.e. the mean-field approximation becomes nearly exact.
\item[(iii)] when $0 < k < \Lambda $, $\widetilde{R}_{k}(q^2)$ is a decreasing function of $q$ which tends rapidly
to $0$   for $\arrowvert q \arrowvert >k $, thus the rapid modes are 
unaffected by the massive term. On the contrary the slow modes have a large mass which decouples them from the fast modes.
\end{itemize}

The physics of the k-systems is encoded in their partition functions 
\begin{equation}
\label{Zk}
 Z_k\left[J \right] =  \int \mathcal{D} \varphi \exp\left( 
-S_{k}\left[ \varphi\right] + J \cdot \varphi
\right)  \; ,
\end{equation}
where  the functional measure $ \mathcal{D} \varphi $ integrates over \textbf{all the field
modes}  $\widetilde{\varphi}_q$ ($\arrowvert q \arrowvert<\Lambda$) and 
$J \cdot \varphi \equiv \int_{x} J_x \, \varphi_x $ ($J$ source). At scale  ``k'' the 
slow modes $\widetilde{\varphi}_q$, $\arrowvert q \arrowvert < k$, 
are frozen by their mass  $\widetilde{R}_{k}(q^2) \sim Z_k  k^2$ and $W_k \left[J \right] = \log  Z_k\left[J \right]$ 
can therefore be interpreted  as  the generator of the connected
Green's function with an IR cut-off. It is a convex function of the source $J$ and it's Legendre transform
defined as
 \begin{equation}
\label{Legendre}
\overline{\Gamma}_{k}\left[ \phi \right] = \sup_{J}\left( J \cdot\phi - W_{k}\left[ J \right]  \right)    \; ,
\end{equation}
is also a convex functional of  the order parameter $\phi = <\varphi>$. $\overline{\Gamma}_{k}\left[ \phi \right]$
is the `` true'' Gibbs free energy of the  k-system. However  it proves convenient 
to introduce and consider rather the  \textbf{ effective average action} of Wetterich
 \begin{equation}
\label{Wett}
\Gamma_{k}\left[ \phi \right] = \overline{\Gamma}_{k}\left[ \phi \right] -\dfrac{1}{2} \phi \cdot \mathcal{R}_{k} \cdot \phi \; .
\end{equation}
By contrast with  $  \overline{\Gamma}_{k}\left[ \phi \right] $    this functional of the field $\phi$ is non-convex but has simple 
limits which follow from the mathematical properties  of the cut-off 
function $ \mathcal{R}_{k}$~\cite{Wetterich}, \textit{i.e. }:
\begin{itemize}
 \item when $k = \Lambda$ no fluctuations has been integrated out.
and we will suppose that $\Gamma_{k=\Lambda}\left[\phi\right] \equiv  S_{\Lambda}\left[ \phi\right]$
(mean field approximation).
 \item when $k = 0 $ all fluctuations have been integrated out and $\Gamma_{k=0}\left[\phi\right] = \Gamma \left[\phi\right]$ is
 the Gibbs free energy of the model under study.
\end{itemize}

Therefore when $k$ decreases from its initial value $\Lambda$ to $k=0$ all the fluctuations of the field are progressively integrated out
and the effective action   $  \Gamma_{k}\left[ \phi \right] $  flows from the bare action $S_{\Lambda}\left[ \phi\right]$ (\textit{i.e.}
 the mean-field approximation for $\Gamma\left[ \phi \right]$)
 to the Gibbs free energy $\Gamma \left[\phi\right]$ of the model. 
The flow equation reads as~\cite{Wetterich}
\begin{equation}
 \label{flow1}
\partial_k \Gamma_{k}\left[ \phi \right] = \frac{1}{2}  \int_{q} \partial_k \widetilde{\mathcal{R}}_{k}(-q,q)\;
\{  \widetilde{\mathcal{R}}_{k} +  \widetilde{\Gamma}_{k}^{(2)}                                
\}^{-1}(-q,q)  \; ,
\end{equation}
where the inverse in the r.h.s. of Eq.\ \eqref{flow1} has to be understood in the operator sense, $ \Gamma_{k}^{(2)}(x,y)=\delta^{2}\Gamma_{k}
[\phi]/\delta \phi_x \delta \phi_y $ is the vertex function of order 2 and $ \widetilde{\Gamma}_{k}^{(2)}(p,q)$ denotes is $2 d$ dimensional
Fourier transform. Note that  $ \Gamma_{k}^{(2)}$ depends functionally
on the field $\phi$ and the  equation for $\Gamma_{k}\left[ \phi \right]$ is thus not closed. One has to resort
to approximations to solve the flow.

\subsection{\label{Thres}The local potential approximation }
A simple, non trivial way to tackle with Eq.\ \eqref{flow1} is to restrict the functional space to functionals
of the form 
\begin{equation}
\label{lpa}
 \mathbf{ (LPA} \, \mathtt{ansatz)} \; \;  \Gamma_{k}\left[ \phi \right]=\int_x \left\{ \frac{1}{2} (\vec{\triangledown} \phi_x )^{2} + U_k(\phi_x) \right\} \; ,
\end{equation}
which constitutes the popular local potential approximation (LPA). Note that in this scheme 
there is no field-renormalization so that $Z_k=1$.
We also stress that the \textbf{ local potential}  $ U_k(\phi_x) $ is a function of
the local  field $\phi_x$ at point $x$, not a functional. Combining Eqs.\ \eqref{flow1} and \eqref{lpa} one then obtains
a closed partial differential equation (PDE) for the  $U_k$ which reads as
\begin{equation}
 \label{flowU1}
\partial_k  U_k(M) =  \frac{1}{2}   \int_{q} \dfrac{\partial_k \widetilde{R}_k(q^2)}{  q^2 + \widetilde{R}_k(q^2) + U_k''(M)} \; ,
\end{equation}
where  the potential $U_k(M)$ is evaluated for a uniform
magnetization  $\phi_x = M$ and  $ U_k^{''}(M) \equiv \partial^2 U_k(M)/ \partial M^2 $ denotes its partial derivative with
respect to $M$. The PDE\ Eq.\ \eqref{flowU1} is non-linear but however quite easy to solve numerically 
above the critical
point  ($r>r_c$,  in the case of the $\varphi^4$ potential). 
Difficulties arise in the ordered phase. To understand why,  we rewrite\ Eq.\ \eqref{flowU1} under a more
convenient form. We define the dimensionless variable $t=-\log( k / \Lambda)$, the RG time chosen to increases from 
0 to $+\infty$ when $k$ decreases from
$\Lambda$ to 0, the dimensionless variable $y=q^2/k^2$ and the dimensionless functions of ``y'' :
$r(y)=\widetilde{R}_k(q^2)/q^2$ (so that $\widetilde{R}_k(q^2)=y \, r(y) k^2$),                      
 $s(y)=-2 y^2 r^{'}(y)$, and $t(y)=y(1+r(y))$ so that 
\begin{equation}
  \label{flowU2}
\partial_t  U_k(M) = -2 v_d k^d \mathcal{L}(U_k''(M)/k^2) \; ,
\end{equation}
where $v_d^{-1}=2^{d+1} \pi^d \Gamma(d/2)$ and the \textbf{threshold function} $ \mathcal{L}$ (according to Wetterich terminology)
is defined as
\begin{equation}
\label{L}
 \mathcal{L} :\omega  \longmapsto \frac{1}{2} \int_{0}^{\infty}\dfrac{y^{\frac{d}{2}-1} s(y) \,  dy}{t(y) + \omega} \; .
\end{equation}

This function $\mathcal{L}(\omega)$ plays a central role in the mathematical analysis of Sec.~\ref{The} and we need first to study
its relevant analytical properties.
For reasonable cut-off functions the two functions  $s(y)$, and $t(y)$ are positive for $y \ge 0$ so that $ \mathcal{L} (\omega)$
is a decreasing function of it's argument. In any reasonable case the function is defined on the interval $]\omega_0, + \infty [$ where 
$\omega_0$ is the largest pole of the  integrand in the r.h.s. of Eq.~\eqref{L} (for any reasonable choices of the cut-off $r(y)$ there 
is  in general only a single pole). Therefore $ \mathcal{L} (\omega)$ decreases from 
$+\infty$ to 0 when  $\omega$ increases from  $\omega_0$ to $ + \infty$.  Finally note that $U_k''(M)/k^2$ is indeed a dimensionless
quantity which gives sense to Eq.\ \eqref{flowU2}. 

Before we tackle the general case let us  consider two important cases. First, the (ultra) sharp cut-off defined by the 
regulator $\widetilde{R}_k(q^2)= \beta  k^2 (1-\theta(q^2 - k^2))$ where  the parameter $\beta \to + \infty $. In that case one shows that
$ \mathcal{L} (\omega) = -\ln (1 + \omega)$ \cite{Wetterich}, so that the flow equation reads as
\begin{equation}
 \partial_t  U_k(M) =  2 v_d k^d \ln (1+ U_k''(M)/k^2) \; .
\end{equation}

A second important case that we shall consider is Litim's regulator  $\widetilde{R}_k(q^2)= ( k^2 -q^2) (1-\theta(q^2 - k^2))$ which yields
$ \mathcal{L} (\omega) = (2/d)/(1 + \omega)$ and  the flow equation \cite{Litim}
\begin{equation}
\label{blu}
 \partial_t  U_k(M) =  -\frac{4 v_d}{d } \frac{k^d}{1+ U_k''(M)/k^2} \; \; .
\end{equation}

\begin{table*}[h!]
\caption{\label{constantes} Constants characterizing the exponential smooth cut-off
threshold function in the case $\alpha=6$. }
\begin{tabular}{|l|r|} 
\hline
$y_0$ = &  2.36806229624554475822396946574                     \\ \hline
$\omega_0$ = & -3.83637249545350290077137812782                   \\ \hline
$K_0$ = &   34.2450749455029367943794121780                \\ \hline
$K_1$ = &   26.7493978693695013104149308428            \\ \hline
$K_2$ = &     -123.598329790226947089342622508          \\ \hline
$K_3$ = &      600.184183291720248551501270440        \\ \hline
$C_1$ = &     26.7493978693695013104149308428         \\ \hline
$C_2$ = &     -4.62060231762294359247562464639         \\ \hline
$C_3$ = &     .0406488093691634129962437962261        \\ \hline
\end{tabular}
\end{table*}

Let us  consider now a regular smooth cut-off function. We have retained the exponential  regulator 
widely  used in recent numerical studies (see \textit{e. g.} ref~\cite{Canet}) 
\begin{equation}
\label{tutu}
\widetilde{R}_k(q^2)= \alpha q^2/(\exp(q^2/k^2) -1)
\end{equation}
that is,  written  in reduced form 
$r(y)=\alpha/(\exp(y) -1)$, where $\alpha$ is some positive parameter. It is easy to see   that
\begin{itemize}
 \item[(i)]  When $\alpha > 2$ the function $t(y)=y(r(y)+1)$, defined for $y \in [0, + \infty [$,
 exhibits a single minimum at some $y_0 >0$ and $t'(y_0)=0$.
 \item[(ii)] When $\alpha = 2$ the minimum of $t(y)$ occurs precisely at  $y_0 = 0$ and $t'(0)=0$
 \item[(iii)]   When $0< \alpha < 2$ the minimum of $t(y)$ on the interval $(0,\infty)$ is located at  $y_0 = 0$ but $t'(0)>0$.
\end{itemize}
The choice $\alpha =6$ yields the better critical exponents according to the authors of reference~\cite{Canet} so we
will retain this value; we are thus in the case (i)  where the minimum of function $t(y)$ is located at some non-zero value
$y_0$. We introduce $\omega_0 = -\min_{ \;0 \leq y \leq \infty} t(y) = -t(y_0)$. 
The threshold function $\omega \mapsto \mathcal{L}(\omega)$
is thus a monotonously decreasing function defined on the interval $]\omega_0,\infty [$. Let us precise
now its asymptotic behavior at each boundary  of the interval.

Wetterich \textit{et al.} \cite{Wetterich} have shown how to obtain the behavior of $\mathcal{L}(\omega)$ for 
$\omega \to \omega_0+$; one expands $t(y)$ around its minimum, \textit{i. e.}
$t(y)=-\omega_0 + t_2 \delta y^2 + \mathcal{O}(\delta y^3)$, where $\delta y = y - y_0$ and  recognize
the fact that, at the leading order,
\begin{eqnarray}
\label{asympto1}
\mathcal{L} &\simeq& \dfrac{1}{2} \int_{-\infty}^{+\infty}\frac{y_0^{d/2-1}s(y_0) }{ \omega -\omega_0 +
t_2 \delta y ^2} \; d\delta y \, , \nonumber \\
&\simeq&  K_0 (\omega - \omega_0)^{-1/2} \, ,
\end{eqnarray}
where
\begin{equation}
 K_0 = \frac{\pi y_0^{d/2 -1} s(y_0)}{2 \sqrt{t_2}} \; ,
\end{equation}
and $t_2= t^{''}(y_0)/2$.

  Note that in Litim's case one has $\omega_0=-1$, $t(y)=1$; therefore the previous analysis breaks down
and $\mathcal{L}(\omega)$ diverges as $\sim (\omega - \omega_0)^{-1}$ rather than as $\sim (\omega - \omega_0)^{-1/2}$.
In the case of the ultra sharp cut-off $\omega_0=-1$ but the function $t(y)$ is not well defined yielding a 
logarithmic divergence of 
$\mathcal{L}(\omega)$ as $\omega \to \omega_0= -1$.

On the other hand an asymptotic behavior of  $\mathcal{L}(\omega)$  for $\omega \to \infty$ is
readily obtained from \eqref{L} and reads
\begin{eqnarray}\label{asympto2}
 \mathcal{L}(\omega)& = &\frac{K_1}{\omega} +  \frac{K_2}{\omega^2} + \frac{K_3}{\omega^3 } + \ldots  \nonumber \\
 K_n &=& \frac{(-1)^{n+1}}{2} \int_0^{\infty}dy \; y^{d/2-1}s(y) t(y)^{n-1} \; .
\end{eqnarray}

Since $ \omega \mapsto L = \mathcal{L}(\omega)$ is bijective from $]\omega_0, \infty [$ to 
$]\infty,0[$ it can be inverted and we denote by $\omega = \mathcal{L}^{-1}$ its inverse. This function 
is defined on the interval $]0, \infty [$ where it decreases from $+ \infty$ to $\omega_0$. 

For $L \to 0$ we infers from \eqref{asympto2} that
\begin{equation}
\label{asympto3}
 \omega(L)\approx \frac{C_1}{L} + C_2 + C_3 L + \ldots  \; ,
\end{equation}
where 
\begin{equation}
 C_1=K_1 \;  \;  ; \;  \; 
 C_2 =  \frac{K_2}{K_1}  \;  ;\; \; 
C_3 = \dfrac{K_3 - K_2^2/K_1}{K_1^2} \; ,
\end{equation}
and for $L \to \infty$ 
\begin{equation}
\label{asympto4}
\omega(L) =  \omega_0 + \frac{K_0^2}{L^2} +\ldots  \; .
\end{equation}
The values of the constants $y_0, \omega_0, K_0, K_1, K_2, K_3, C_1, C_2, C_3$ are determined numerically; they are resumed
in Table~\ref{constantes} in the case $\alpha=6$.
For numerical applications the functions $\omega(L)$ or $\mathcal{L}(\omega)$, as well as their derivatives if required,
have been fitted by polynomial expressions taking into account their asymptotic behaviors.
\subsection{\label{The}Flow equations for the threshold functions}
\begin{figure}[t!]
\includegraphics[angle=0,scale=0.55]{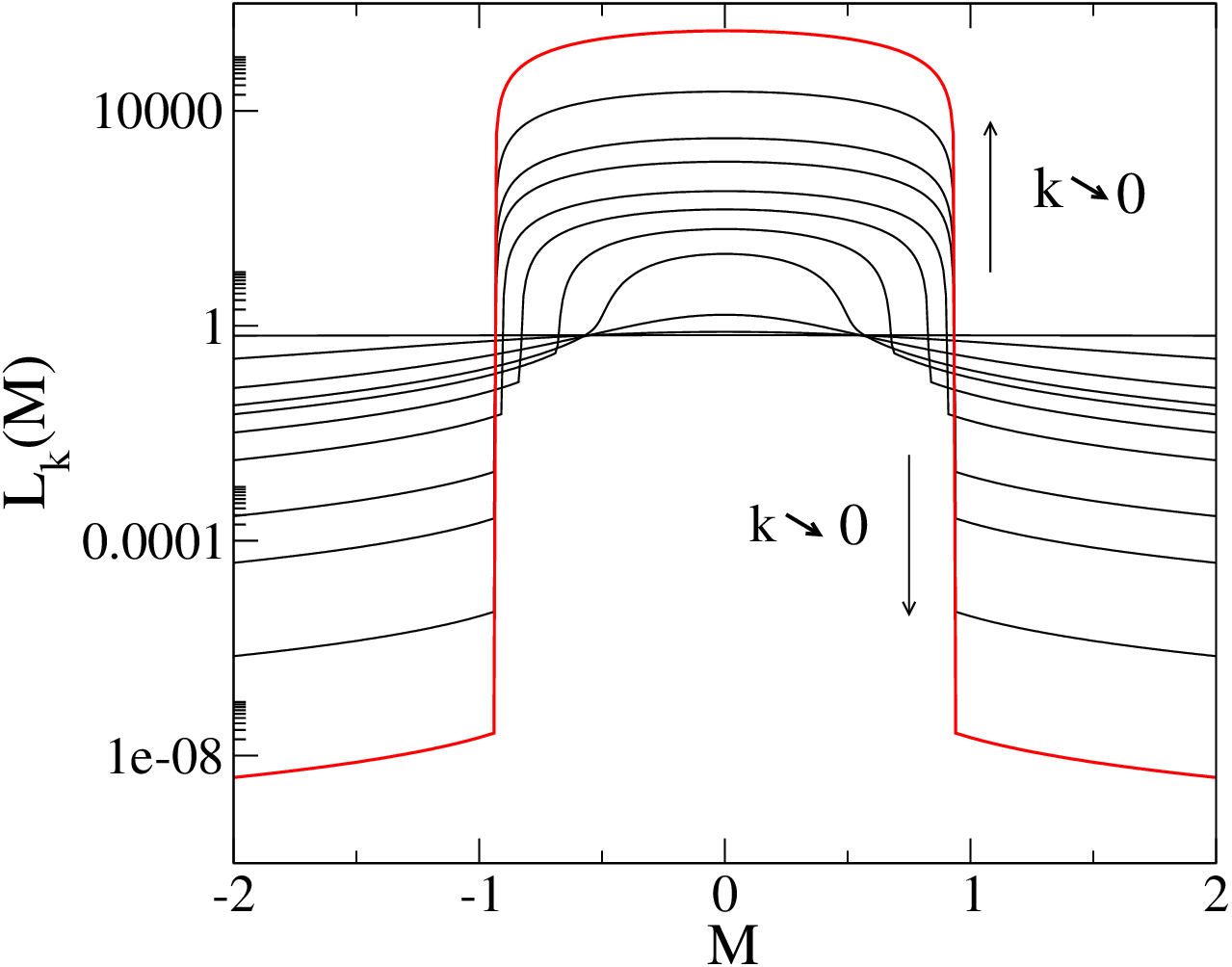}
\caption{
\label{lnL} Litim's approximation : log-plot of the threshold function $L_k(M)$
 for different value of the scale $k_{\textrm{min}}\leq k \leq \Lambda$. The $\varphi^4$ potential
 is used  at initial scale  $\Lambda=10$ with $g=1.2$ and $r=-0.35$, $M_\textrm{max}=8$. Red curve
: $k_{\textrm{min}}=10^{-6}$.
}
\end{figure}
As suggested first by  Bonanno and  Lacagnina~\cite{Bonanno} (however in a more restricted context)
it is convenient to perform the change of variables
$(M,k,U) \Rightarrow (M,k,L)$ with
\begin{equation}
 L_k(M) = \mathcal{L}(U''_k(M)/k^2) \; .
\end{equation}
The one to one mapping $\omega \Leftrightarrow \mathcal{L}$ insures the mathematical
equivalence of solving the RG flow equations either for $U''_k$ or for $L_k$. 
However, for  numerical reasons 
the RG flow equation for $ L_k$ is much easier to solve than that for $U_k$~\cite{Bonanno}.
The RG flow equation for $L_k$ is readily
deduced from that for $U_k$ (\textit{cf.} \eqref{flowU2}) :
\begin{equation}
 \label{flowL}
L''_k = - \frac{k^{2-d}}{2 v_d} \omega'(L_k)  \frac{\partial L_k}{\partial t} +  \frac{k^{2-d}}{ v_d} \omega(L_k) \; .
\end{equation}
Eq.~\eqref{flowL} is a quasi-linear parabolic  PDE which can be studied analytically in 
some depth, notably in the two phase region,
and for which, in addition, the mathematicians have provided us with robust and efficient solvers in view of a numerical study.
For the exponential smooth cut-off~\eqref{tutu} the function $ \omega(L_k)$ and its derivatives
which enter Eq.~\eqref{flowL} must be determined numerically. It is worthy to write down explicitely these flow equations 
\begin{itemize}
 \item for the sharp cut-off regulator ($\mathcal{L}(\omega)=-\ln(1+\omega)$ and $\omega(L)=\exp(-L) -1$)
   \begin{equation}
  L''_k =  \frac{k^{2-d}}{2 v_d} \exp(-L_k)  \frac{\partial L_k}{\partial t} +  \frac{k^{2-d}}{ v_d} (\exp(-L_k)-1) \; .  
   \end{equation}
\item and for Litim's regulator ($\mathcal{L}(\omega)=(2/d)(1+\omega)$ and  $\omega(L)=2/(d L_k) -1$)
     \begin{equation}
  L''_k =  \frac{k^{2-d}}{ v_d} \dfrac{1}{d L_k^2}  \frac{\partial L_k}{\partial t} +  \frac{k^{2-d}}{ v_d} 
(\dfrac{2}{d L_k} -1) \; .
   \end{equation}
\end{itemize}

\begin{figure}[t!]
\includegraphics[angle=0,scale=0.50]{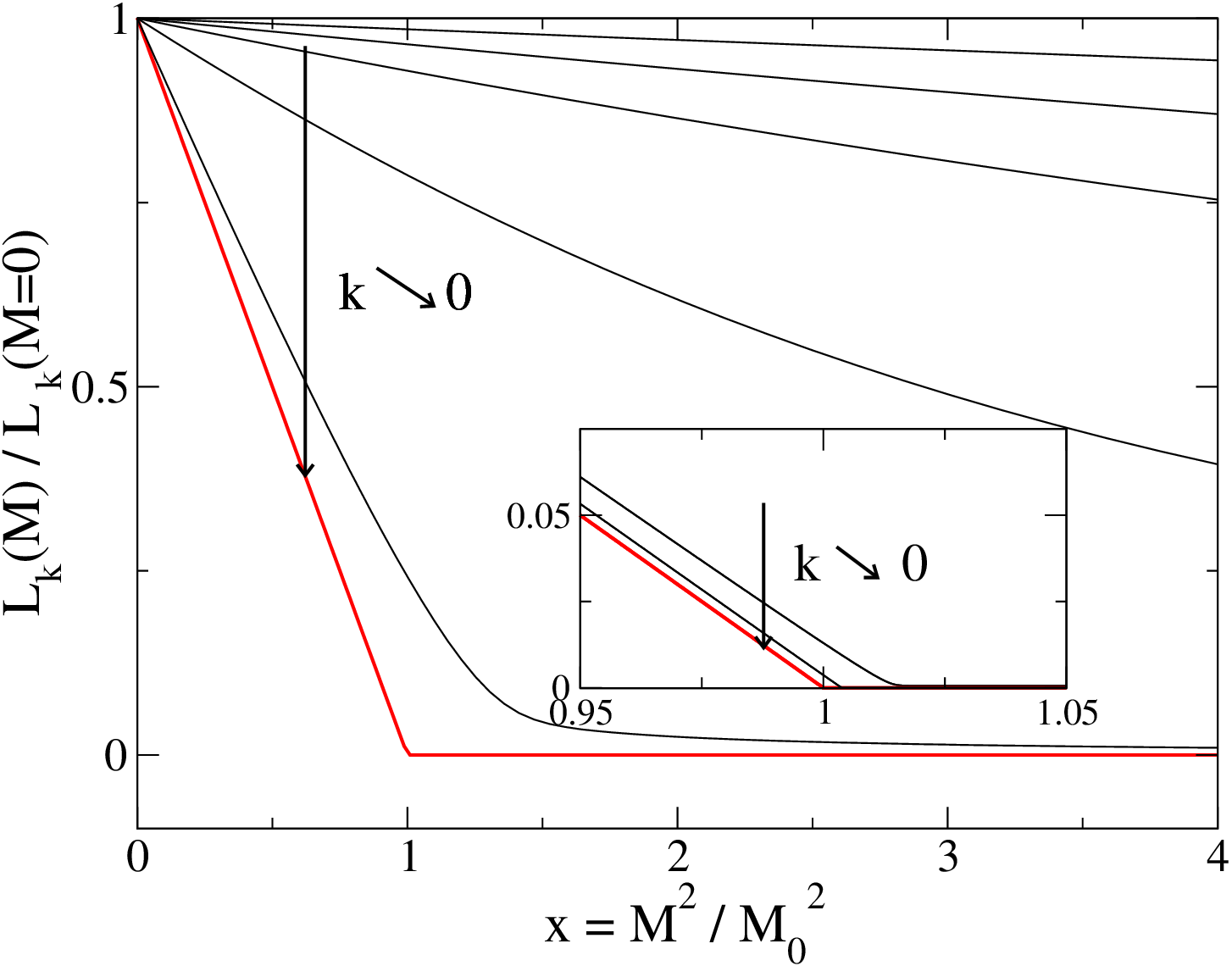}
\caption{
\label{Luniv} Litim's approximation : universal function  $L_k(M)/L_k(M = 0)$
as
a function of $x = M^2/M(0)^2$
for different value of the scale $k_{\textrm{min}}<k<\Lambda$.
Universal curve~\eqref{universal} : red line ( $k=k_{\textrm{min}}=10^{-6}$). The $\varphi^4$
potential is used  at scale  $\Lambda=10$ with $g=1.2$ and $r=-0.35$, $M_\textrm{max}=8$.
}
\end{figure}

\subsection{\label{LPA}Behavior of the flow in the ordered phase}   
\begin{figure*}[!]
\includegraphics[angle=0,scale=0.75]{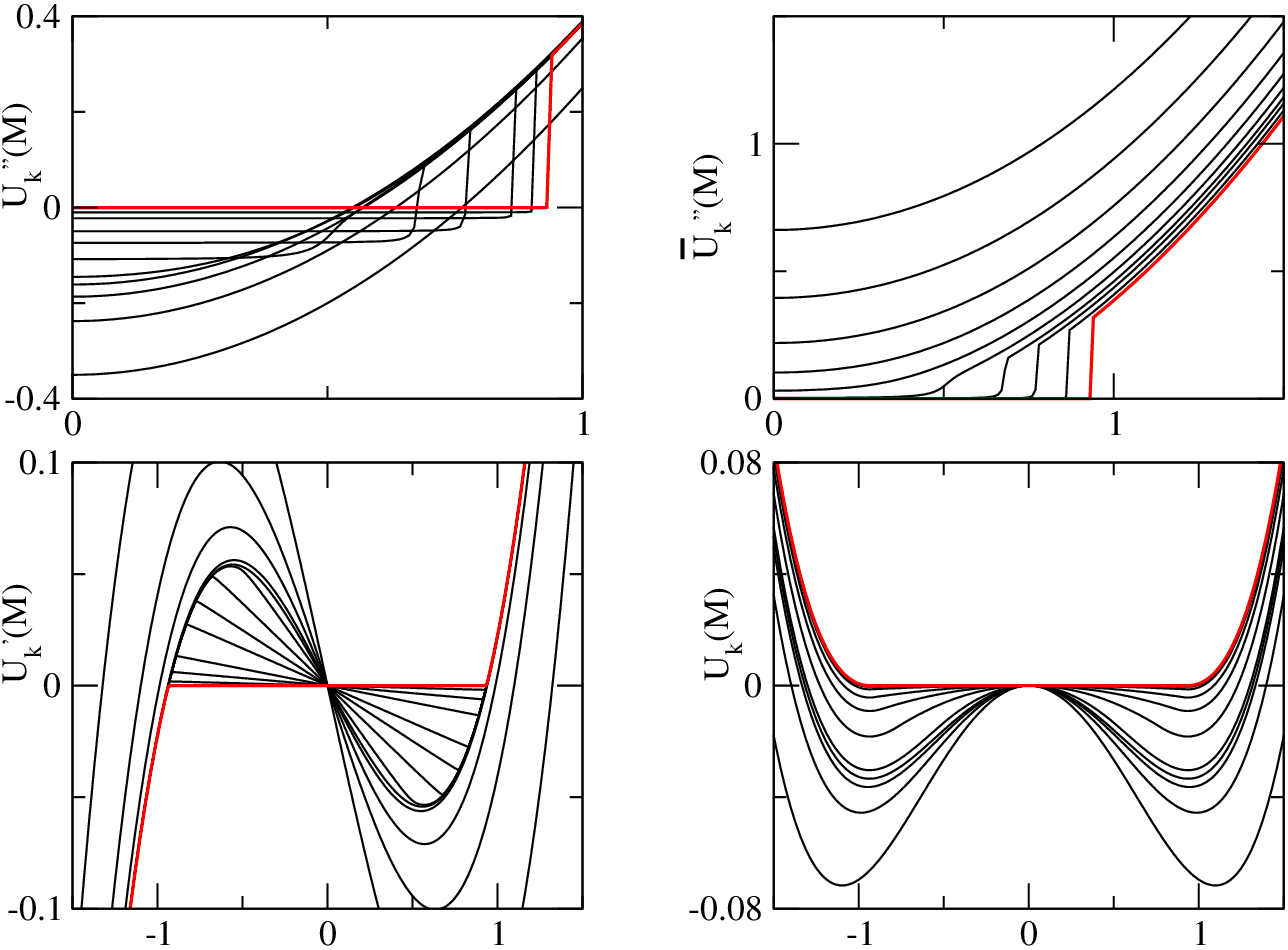}
\caption{
\label{figU} Litim's approximation : from right  to left  and bottom to top : local potential $U_k(M)$, 
 the derivatives  $U_k^{'}(M)$ (magnetic field) , $\overline{U}_k^{''}(M)$ (inverse susceptibility), and $U_k^{''}(M)$ 
for different values of $k$ in the range $k_{\textrm{min}}<k<\Lambda$, with $k_{\textrm{min}}=10^{-6}$.
The $\varphi^4$  potential is used at scale  $\Lambda=10$ with $g=1.2$, and $r=-0.35$ ($M_\textit{max}=8$ ).
 Red lines :  result at  $k =k_{\textrm{min}}$.
}
\end{figure*}

Below $T_c$,  a spontaneous magnetization $M_0(r)$ can settle
in the system in the absence of an external magnetic field $J$. More precisely, 
in the thermodynamic limit, any magnetization $-M_0(r) \leq M \leq M_0(r)$
is likely to settle with the same probability. Roughly speaking we have a coexistence region of two (or more)
magnetized phases $\pm M_0$ analogous to a liquid-vapor coexistence. Therefore 
$\Gamma''(M)=\partial J/\partial M \propto \widetilde{\Gamma}^{(2)}(0)=0$ from which it follows
that, in the limit $k \to 0$ the denominator of the r.h.s. of Eq.~\eqref{flowU1} tends to zero.
Therefore the threshold function  $L_k(M)$
should become large and positive as $k\to 0$  and finally it should diverge to $+\infty$ at $k=0$ for
any magnetization in the two phase region $-M_0 \leq M \leq M_0$.

With the hypothesis that $L_k$ is large and positive, $\omega(L_k) \sim \omega_0$ as follows
from~\eqref{asympto4}; then the flow equation~\eqref{flowL} for $L_k$ simplifies to
  \begin{equation}
\label{zou}
    L''_k = \frac{k^{2-d}}{v_d} \omega_0 \; ,
  \end{equation}
which can be integrated to give
 \begin{equation}
\label{L_univ}
  L_k(M) = \frac{-k^{2-d}}{2 v_d}\;  \omega_0 \, (M_0(k) ^2 - M^2)  \; .
 \end{equation}
 for   $ -M_0(k) \leq M \leq M_0(k) $ where $M_0(k)$  is the  constant of integration (depending on $k$)
of Eq.~\eqref{zou}.
Clearly $M_0(k)$ can be interpreted as a precursor of the magnetization of the system at scale $k$.
Note that  $M_0(k)$  can be obtained numerically from the numerical solution of PDE~\eqref{flowL}
from the value of $L_k(M=0) = -\omega_0 M_0(k)^2k^{2-d}/(2 v_d)$  at the origin.
As expected one finds that for $d>2$ the function $ L_k(M) $ diverges to $+ \infty$ (recall
that  $\omega_0 < 0$)  when $k \to 0$ as $\propto k^{2-d}$ for any magnetization $M$ in the two phase region.

Outside the coexistence region -\textit{i.e.} for $\arrowvert M \arrowvert > M_0(k)$- one expects a finite compressibility
and thus $L_k(M) $ tends to $0$ as  $ \sim K_1 k^2/U''_k(M) $ with $k$, as follows from \eqref{asympto2}. This behavior of  $ L_k(M)$ is
exemplified in figure~\ref{lnL} (details concerning the numerical procedure to solve \eqref{flowL} will be given in Sec. III).

One of the consequences of Eq.~\eqref{L_univ} is the universal behavior :
\begin{align}
\label{universal}
   L_k(M)/L_k(0) &= 1- x  \text{ with } x= (M/M_0(k)) ^2  \; \; \text{ for } \mid x  \mid \leq 1 \nonumber \\
 L_k(M)/L_k(0)  & = 0 \; \; \text{ for }  \mid x  \mid \geq 1   \; .
\end{align}
The function $L_k(M)/L_k(0) $ of the argument $x=(M/M_0(k)) ^2 $   is thus universal as $k \to 0$ , \textit{i.e.} it is 
independent of the thermodynamic state (provided the temperature satisfies $r<r_c$), the regulator $R_k(q^2)$,
 the value of $\omega_0$, and
the dimension of space $d > 2$. This remarkable features are illustrated in figure~\ref{Luniv}.

One can of course infer the behavior of the potential $U_k(M)$ and its derivatives from that of $L_k$ and one has,
for $ -M_0(k) \leq M \leq M_0(k)$
\begin{eqnarray}
\label{U}
 U^{''}_k(M) &=& k^2 \omega_0 \nonumber \\
 U^{'}_k(M) &=& k^2 \omega_0 M \nonumber \\
 U_k(M) &=& \frac{1}{2}k^2 \omega_0 M^2 + \text{constant} \; ,
\end{eqnarray}
where we have noted that $ U'_k(M=0)=0$ for a $Z_2$ symmetry; see figure~\ref{figU} for an illustration.
In the special case of  the sharp cut-off and Litim regulator recall that one has to set $\omega_0 = -1$ in the above equations. 
Another interesting consequence of this simple behavior for the potential and its derivatives  is the expression of the
''true''  susceptibility of the k-system (cf Eq.~\eqref{Wett}) 
which reads as
\begin{eqnarray}
\label{Ubar}
 \dfrac{d^2}{d M^2} \overline{\Gamma}_{k}(M) &=& \overline{U}^{''}_{k}(M) \nonumber \\
&=& U^{''}_k(M)+  \left .y r(y) \right\vert_{y \to 0} \nonumber \\
&=& k^2 (t(0) -\min_{0 \leq y \leq \infty}t(y) ) \; ,
\end{eqnarray}
and is thus  positive  at any $k \ne 0$. However the strict convexity of $\overline{\Gamma}_{k}(M)$ is  preserved at each
step of the RG flow in the LPA approximation only  if $\alpha >2$ as follows from the discussion after Eq.~\eqref{blu}. Note that
 in that case Eq.~\eqref{Ubar} becomes $\dfrac{d^2}{d M^2} \overline{\Gamma}_{k}(M)
=(\alpha + \omega_0) k^2$. In the case $\alpha \leq 2$, strict convexity could be violated and the validity
of the LPA is no more guaranteed.

\begin{figure}[!]
\includegraphics[angle=0,scale=0.55]{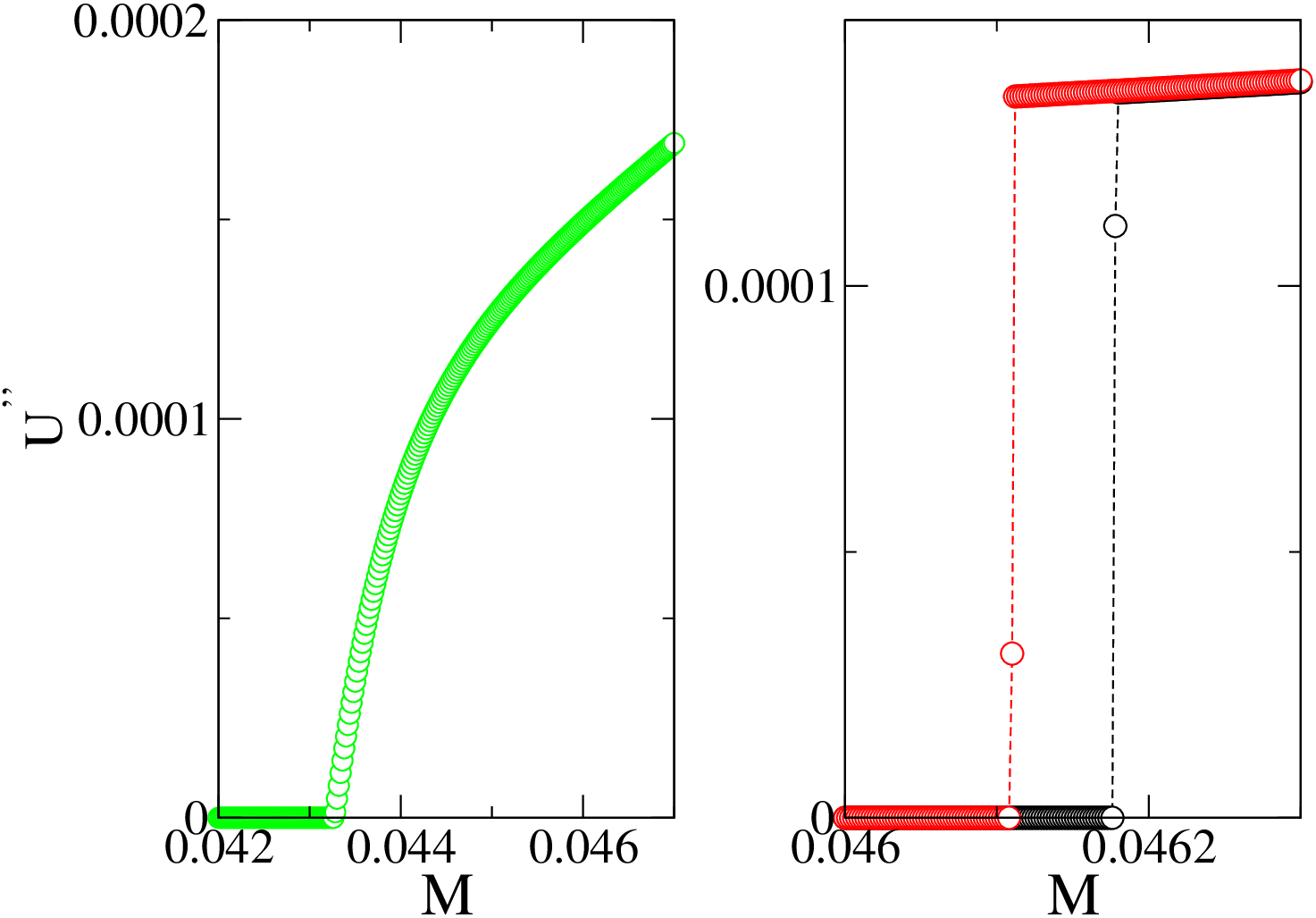}
\caption{
\label{LS}  $U_{k_{\text{min}}}^{''}(M) $ with sharp  (green solid circles), Litim's (black solid circles)
and exponential smooth cut-off (red solid circles) in $d=3$ dimensions of space.
 In all cases : $\varphi^4$ potential at $\Lambda =10$ with $g=1.2$, $M_{\text{max}}=8$,   $\delta M = 2. 10^{-6}$,
and $k_\text{min}= 10^{-8}$.
Litim's cut-off : $r=-0.197032$, sharp cut-off  : $r=-0.297286$, exponential smooth cut-off : $r=-0.412195$.
}
\end{figure}

The simple behavior described by Eqs.~\eqref{U} is independent of the regulator and has
been known from long  \cite{Reatto0,Bonanno}.
However Parolla \textit{et al.} \cite{ReattoI,ReattoII,Reatto0} 
were the first to remark, in a more refined discussion, that the behavior
of the susceptibility for $M \to \pm M_0(k)$ depends strongly on the asymptotic behavior
 of $\mathcal{L}(\omega)$ for $\omega \to \omega_0 +$.
For the sharp cut-off regulator they found that  $U^{''}_k(M)$ is continuous and equal to zero at 
 $M=M_0$  (for $d<4$) (see Ref.~\cite{ReattoI}) while for the Litim regulator
 they found that, for $d>2$,  $U^{''}_k(M)$  exhibits a discontinuity from a
 finite positive value at $M=M_0+$ to a  zero value  at $M=M_0 -$
( cf Ref.~\cite{ReattoII}). This discontinuity of
the susceptibility $\chi $ jumping from an infinite value in the two phase region to a finite one outside, 
is of course the expected physical behavior for $\chi $.
In other words, in the case of the sharp cut-off regulator,  the spinodal and the binodal curves merge in a single curve, which is a serious flaw in the theory
(see figure~\ref{LS}).

We now extend the analysis of Parolla \textit{et al.} to the quite general case :
\begin{equation}
\label{simpl}
 \mathcal{L}(\omega) = \dfrac{K_0}{(\omega - \omega_0)^{\nu}} \; ,
\end{equation}
where $\nu >0$ is an arbitrary exponent, not to be confused with the critical exponent of the correlation length!
This expression includes notably the case of the exponential smooth cut-off considered in 
this paper - with $\nu = 1/2$, cf Eq.~\eqref{asympto1}- since
only the singularity of $\mathcal{L}(\omega)$ as $ \omega \to \omega_0 +$ is relevant to study the behavior
of $U^{''}_k(M)$ as $M \to \pm M_0$.
As follows from~\eqref{L_univ}, when $k\to 0$ $L_k(M)$, diverges therefore  to $+ \infty$ for $\vert M \vert < M_0$ for $d>2$ 
and tends uniformly to $0$ outside the interval $]-M_0, M_0[$ (see fig.~\ref{lnL}). 
This behavior is too ``drastic'' to discriminate the analytical properties 
of $U^{''}(M)$ in  the vicinity of $\pm M_0$. Borrowing an idea of Reatto \textit{et al.} we introduce the following new function of the field
\begin{equation}
 \label{F} F (k,M) = \frac{1}{(U^{''}_k(M)-k^2 \omega_0)^{\nu}} = K_0^{-1}k^{-2 \nu} L_k(M) \; .
\end{equation}
In the two  phase region  $U^{''}_k(M) \to 0$ as $k \to 0$ and thus $F \to + \infty$, while outside  $0 < U^{''}_{k=0}(M) <\infty$ yielding
$F$ to reach a finite value $F_0$ at $k=0$.
It follows from~\eqref{L_univ} and~\eqref{F} that for $k \to 0$ and  $ \vert M \vert  \leq M_0(k)$ 
\begin{eqnarray}
 F(k,M) &=& k^{2-d -2 \nu} \omega_0 (x^2 -x_0(k)^2) \nonumber \\
 x  &=& \frac{M}{\sqrt{2 v_d K_0}} \; , \nonumber \\
 x_0(k)  &=& \frac{M_0(k)}{\sqrt{2 v_d K_0}} \; .
\end{eqnarray}
We are led to introduce the new variable
\begin{equation}
 z = (x -x_0(k)) k^{2-d -2 \nu}
\end{equation}
such that, at fixed $z$ and for $k \to 0$, one has
\begin{equation}
\label{asympt}
 F(k,M) \approx 2 \omega_0 x_0(k) z \; ,
\end{equation}
valid if $2 \nu >-d+2$, \textit{i. e. } $\nu >0$ if $d>2$. As quoted pleasingly by Parola \textit{et al.}\cite{ReattoI,ReattoII}
this variable $z$ allows us to ``zoom`` in the region $x \approx x_0(k)$. Inside the two-phase region
$z \to - \infty$ and we have the asymptotic behavior~\eqref{asympt} of $F$, while for
$z \to  \infty$, \textit{i. e.} outside the the two-phase region,  the function $F$ should reach a finite value $F_0$.
To avoid a proliferation of notations we still write $F(k,z)$ the function $F(k,M)$ expressed in terms of its new
variables $(k,z)$. 

The dependence of the reduced spontaneous magnetization
$x_0(k)$ on $k$ as $k \to 0$ will prove of great importance in our analysis. A priori it should be reasonable to assume a k-dependence as
\begin{equation}
\label{a_par}
M_0(k)\sim M_0 - a k^2 \textrm{ with } a>0
\end{equation}

for sufficiently small $k$'s, since $M_0(k)$ should be an analytical function of vector $\vec{k}$. 
Parameter $a>0$  describes
the displacement of the precursor of the spontaneous magnetization $M_0(k) = \sqrt{2 v_d K_0} x_0(k)$
with  scale $k$. With this reasonable assumption we can  finally wrote the RG flow equation :
\begin{widetext}
\begin{eqnarray}
\label{Fond}
z &=& k^{2 - d - 2 \nu} _; (x -x_0 + a k^2) \; ,   \nonumber \\
 k^{4-d-2 \nu} \frac{\partial^2}{\partial z^2}F(k,z) &=& 2 k^2 \omega_0 +
                 \frac{-1}{\nu}\dfrac{1}{F^{1/\nu +1}(k,z))} \times \nonumber \\
&\times& [ k\partial_k F(k,z) + (2-d -2 \nu) z \partial_z  F(k,z) 
+ 2 a  k^{4-d-2 \nu} \partial_z F(k,z) ] \; .
\end{eqnarray}
\end{widetext}

We now analyse the asymptotic behavior of the stationary solutions ($k\partial_k F(k,z)= 0 $)
of  EDP~\eqref{Fond} at $z \to \pm \infty$ in the
limit $k \to 0$.  This is justified since below $r_c$ the flow stops at some finite $k$. These solutions are similar
to scaling solutions at a fixed point; however $F(k,z)$ and $z$ have dimensions and we cannot thus speak 
of scaling solutions \textit{stricto sensu}.
Obviously, in order to find these  solutions, which we cannot refrain to christen as fixed point solutions despite
our previous remarks,
we have to discriminate our study according the sign of the exponent  $4 -d -2 \nu$ in Eq.~\eqref{Fond}.
\subsubsection{\label{I} $4 -d -2 \nu < 0$}

\begin{figure}[!]
\includegraphics[angle=0,scale=0.65]{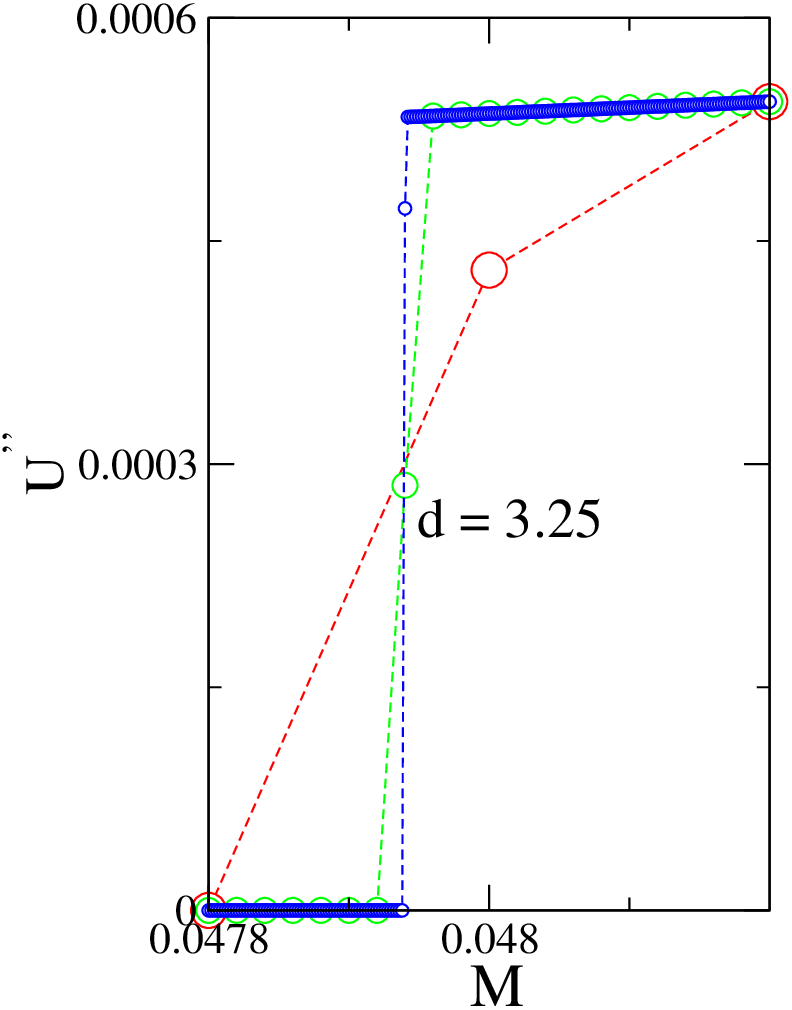}
\caption{
\label{LS325} 
$U_{k_{\textrm{min}}}^{''}( M ) $ for the simplified smooth cut-off (\textit{i. e.} $\nu=1/2$ in
 Eq.~\eqref{simpl}) in the vicinity of $M_0$.
$\varphi^4 $ model with $\Lambda=10$, $g=1.2$, $k_{\textrm{min}}=10^{-8}$, 
$M_{\textrm{Max}}=8$, 
$d=3.25$ and $r= -0.1545$.
Color code for the open circles (if present) : black  :  $\delta M = 2. 10^{-3} $, red : $\delta M = 2. 10^{-4}$, green :  $\delta M = 2. 10^{-5}$,
blue : $\delta M = 2. 10^{-6}$.
All the points of the grid of field values are reported, dashed lines are only guide-lines for the eyes.
}
\end{figure}

Neglecting terms which tend to 0 when $k \to 0$ in Eq.~\eqref{Fond} one finds the implicit
fixed-point equation :
\begin{equation}
 \frac{\partial^2}{\partial z^2} F(z) = 2 a \frac{-1}{\nu}\frac{1}{F^{1/\nu + 1}(z)} \frac{\partial}{\partial z} F(z), 
\end{equation}
where $F(z) \equiv F(k=0,z)$.
This is an autonomous differential equation which can be solved analytically :
\begin{eqnarray}
 \label{soluI}
-c z + d &=& F_0  \; \mathcal{F}(F/F_0) \; ,
\end{eqnarray}
where $c>0, d$ are integration constants, $F_0 = (2a/c)^{\nu}$ and $\mathcal{F}(X)$ a primitive of
$f(X) = 1/(1 - X^{-1/\nu})$. We define $n_0$ as the integer part of $\nu$, $\delta$ as its decimal part, \textit{i. e. }
$\nu = n_0 + \delta$, and we have 
\begin{eqnarray}
  \label{soluII}
F(X) &= & X + \dfrac{\nu}{\delta}\; (X^{\delta/\nu}-1) + \nu \; 
(1 - \delta_{n_0,1}) \sum_{n=1}^{n_0 -1}\dfrac{X^{(\nu - n)/\nu}}{\nu-n}  \nonumber \\
  &-& \nu X^{(\delta-1)/\nu} \Phi(X ^{-1/\nu},1, 1-\delta) \; ,
\end{eqnarray}
where the third term in the r.h.s. of the equation is present only if $\nu\geq2$ and $\Phi(z,s,a)$
is the Lerch transcendent \cite{grad,Lerch}. This curiosity is defined on the disk $\vert z \vert <1$ (although it can be extended
by analytical continuation to the cut plane   $\mathbb{C} - [1,\infty [ $). It is defined as
\begin{eqnarray}
 \label{Lerch}
\Phi(z,s,a) &=& \sum_{n=0}^{\infty}\dfrac{z^n}{(a + n)^s} \; , \nonumber \\
&&a \ne 0, -1, \ldots , \; \; \vert z \vert <1 ; \vert z \vert =1 , \Re(s) >1
\end{eqnarray}

When $\nu = n_0$ is an integer Eq.~\eqref{soluII} reduces to the simple form :
\begin{eqnarray}
  \label{soluIII}
F(X) &= & X + \ln X  + n_0  \; 
(1 - \delta_{n_0,1}) \sum_{p=1}^{n_0 -1}\dfrac{X^{p/n_0}}{p} \nonumber \\
&+& n_0 \ln(1-X^{-1/n_0}) \; .
\end{eqnarray}
This result was first obtain in Ref.\cite{ReattoII}   in the case $n_0=1$ (Litim case).
From the known behavior of Lerch function, in particular $\Phi(z^{-1/\nu},1, 1- \delta) \sim - \ln(1- z^{-1/\nu})$ for
$z \to 1+$ one deduces that
\begin{eqnarray}
 \mathcal{F}(X) &\sim& X \text{ for } X\to +\infty \nonumber \\
  \mathcal{F}(X) &\sim & \nu \ln(X^{1/\nu} -1)  \text{ for } X\to 1+ 
\end{eqnarray}

We are now in position to discuss the property of the fixed point solution~Eq.~\eqref{soluI}. In the two phase region, $z \to -\infty$, the function
$F \sim 2 \omega_0 x_0 z$ diverges to $+\infty$, thus $\mathcal{F}(F/F_0) \sim F/F_0$, and one has asymptotically
$ -c z \sim 2 \omega_0 x_0 z$ which fixes the value of the integration constant  $c= -2 \omega_0 x_0$, from which we deduces
$F_0 = (-a/(\omega_0 x_0))^\nu$. In the one phase sector,  $z \to +\infty$ we find that $F \to F_0-$ 
reaches a finite value as $k \to 0$. Returning to our old variables $(k,M)$ we have shown that the fixed
point solution satisfies :
\begin{eqnarray}
\lim_{M \to M_0+} \lim_{k \to 0} F(k,M)& = & 0<  F_0 < \infty     \; , \nonumber \\
\lim_{M \to M_0-} \lim_{k \to 0} F(k,M)& = & + \infty    \; ,
\end{eqnarray}
from which follows that  for $4 -d -2 \nu < 0$ (and $d>2$) the susceptibility $\chi = F(0,M)^{\nu}$ is discontinuous
at $M= \pm M_0$. See Fig.\eqref{LS325} for an illustration.
\subsubsection{\label{II} $4 -d -2 \nu =0$}
\begin{figure}[!]
\includegraphics[angle=0,scale=0.65]{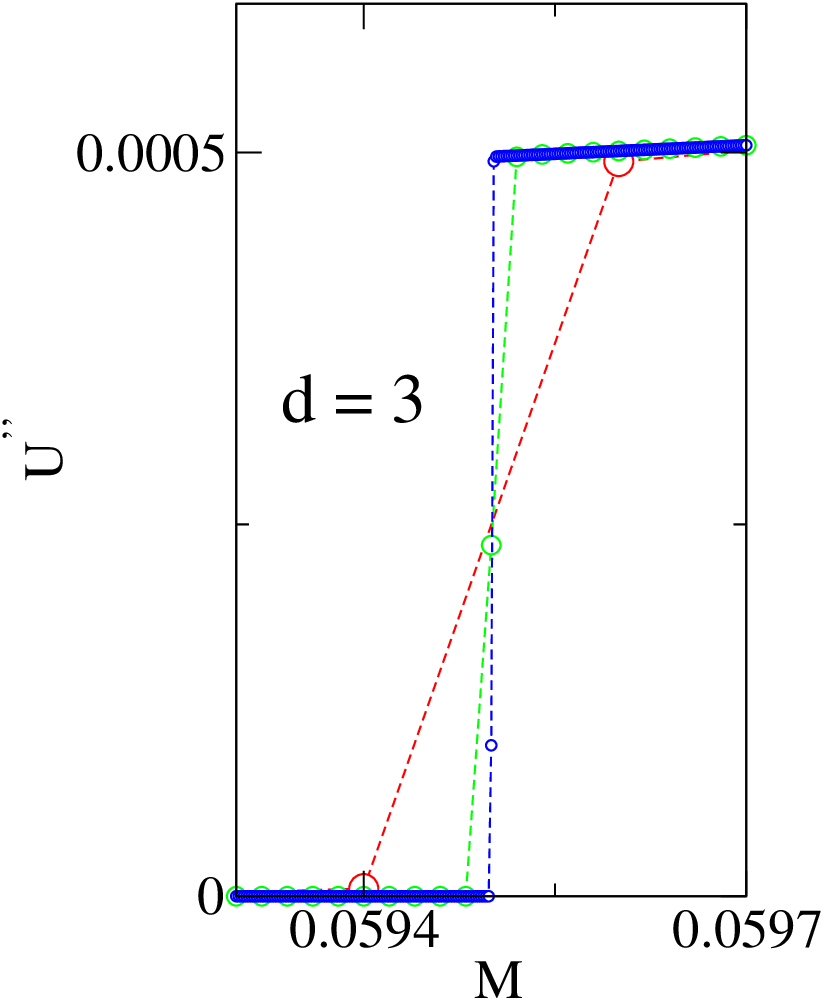}
\caption{
\label{LS3} 
Same  as in fig.~\ref{LS325} except that  $d=3.$ and $r= -0.1495 $.
}
\end{figure}
We consider here the marginal case $4 -d -2 \nu =0$. It is an important one since it corresponds to the general
smooth cut-off ($\nu=1/2$,  cf Eq.~\eqref{asympto1})  in $d=3$. 
Neglecting terms which tend to 0 when $k \to 0$ in Eq.~\eqref{Fond} one finds the implicit
fixed-point equation :
\begin{equation}
 \frac{\partial^2}{\partial z^2} F(z) + 2 (a -z)  \frac{1}{\nu}\frac{1}{F^{1/\nu + 1}(z)} \frac{\partial}{\partial z} F(z) =0
\end{equation}
This equation does not seem to  have an  analytical solution but, asymptotically :
\begin{subequations}
\begin{eqnarray}
 F(z) & \sim &  2 \omega_0 x_0 z \text{ for } z \to -\infty   \; ,  \label{a} 
\\
F(z)  & \sim &  F_0< +  \infty  \text{ for } z \to +\infty \; .  \label{b}
\end{eqnarray}
\end{subequations}
Equation~\eqref{a} is required by the correct matching with the solution~\eqref{asympt} in the two-phase region
while~\eqref{b} guarantees finite susceptibility outside the magnetization curve.  We conclude again in favor of
a discontinuity of the susceptibility at $M=M_0$. See Fig.\eqref{LS3} for an illustration.
\subsubsection{\label{III} $4 -d -2 \nu > 0$}
\begin{figure}[!]
\includegraphics[angle=0,scale=0.65]{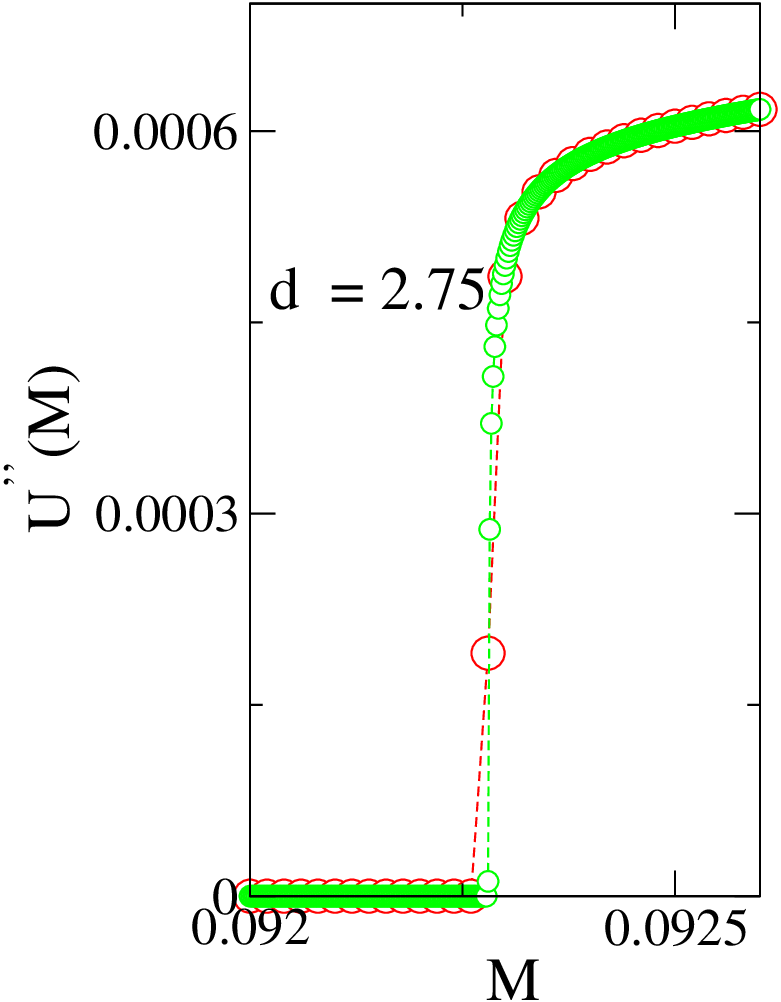}
\caption{
\label{LS275} 
Same   as in fig.~\ref{LS325} except that  $d=2.75$ and $r= -0.1495$.
}
\end{figure}

In this case the term in ''a''  which accounts for the displacement of the spontaneous magnetization 
$M_0(k)$ with scale $k$ (cf Eq.~\eqref{a_par}) can be discarded from Eq.~\eqref{Fond}. The function $F(k,z)$ is no more adapted to our discussion
and we are led to a new change of variables to eliminate the relevant dependence on $k$ from the RG flow equation. Let us introduce
\begin{eqnarray}
\label{G}
 G(k,z) & = & k^{-\mu} F(k,z) \; , \nonumber \\
     \mu & = &  \dfrac{d-4 -2 \nu}{\nu +1} \nu  \;  \; >0  \; .
\end{eqnarray}
By use of these variables the RG flow equation for $G$ admits in the limit $k \to 0$ a stationnary solution $G(z)$  which satisfies 
the following differential equation :
\begin{equation}
 \frac{\partial^2}{\partial z^2} G(z) +   \frac{1}{\nu}\frac{1}{G^{1/\nu + 1}(z)}
\{ \mu G(z) + (2-d -2 \nu)  z \frac{\partial}{\partial z} G(z)\} = 0 \; .
\end{equation}
This equation cannot  be solved analytically but it satisfies the boundary conditions 
\begin{subequations}
\begin{eqnarray}
 G(z) & \propto &  z \text{ for } z \to -\infty   \; ,  \label{aa} 
\\
G(z)  & \sim &    z ^{\mu /(d -2 -2 \nu)} \text{ for } z \to +\infty \; .  \label{bb}
\end{eqnarray}
\end{subequations}
Equation~\eqref{aa} is required by the correct matching with the solution~\eqref{asympt} in the two-phase region,
 while~\eqref{bb} implies a non-physical divergence of the susceptibility on 
the magnetization curve. Indeed it follows from~\eqref{bb}
and~\eqref{G} that $U^{''}_{k=0}=F^{-1/\nu} \propto (M -M_0)^{\varpi}$ ($M>M_0$ since $z \to + \infty)$)  where the exponent 
\begin{equation}
 \varpi = \dfrac{4 - d+ 2\nu}{(\nu+1)(d-2 +2 \nu)} 
\end{equation}
is strictly positive if $d>2$. Not unexpectedly,  one recovers,  in the limit $\nu \to 0$, the exponent $\varpi=(4 - d)/(d-2)$
obtained by Parolla \textit{et al.} in the case of the sharp cut-off \cite{ReattoI}. For the cut-off considered in this section the LPA seems to detect the spinodal  rather than the coexistence curve and  the exponent $\varpi$ can be interpreted as the exponent of the inverse compressibility $\chi^{-1}(M)$ on the spinodal. Note that, however, it depends strongly on the non physical exponent $\nu$
which characterizes the divergence of the threshold function as $\omega \to \omega_0$.

\begin{figure}[!]
\includegraphics[angle=0,scale=0.65]{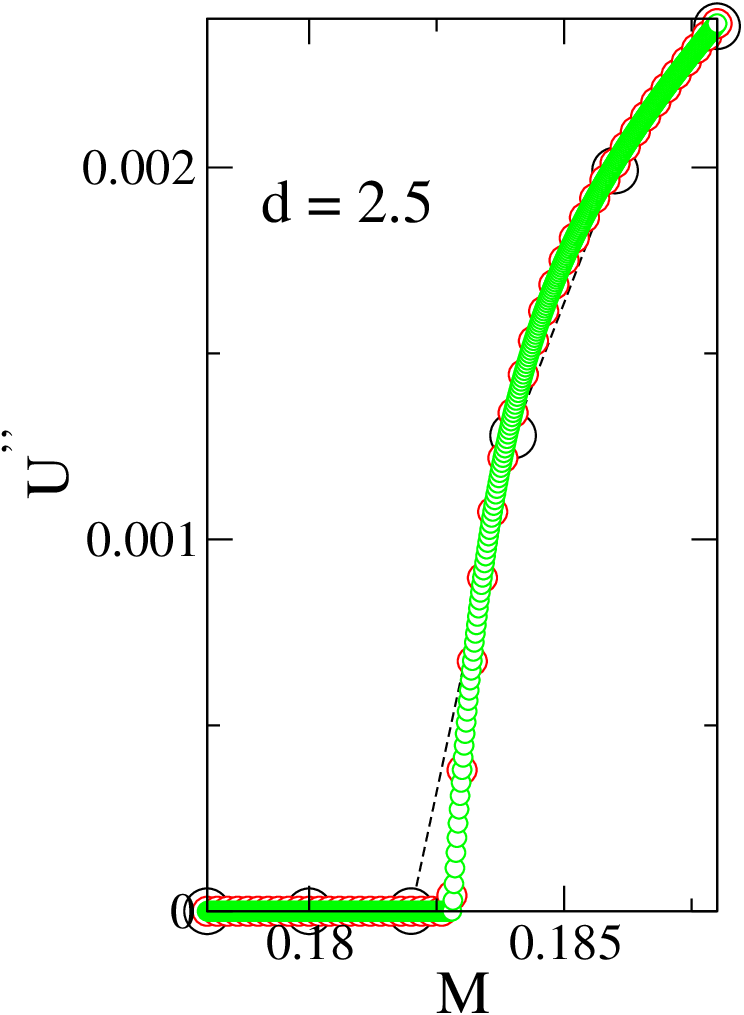}
\caption{
\label{LS25} 
Same  as in fig.~\ref{LS325} except that  $d=2.5$ and $r= -0.157 $.
}
\end{figure}
We conclude that for $4 -d -2 \nu >0$ (and $\nu \geq 0 $ and $d>2$) the LPA has the unphysical feature to give
an infinite susceptibility at $M= \pm M_0$. The isotherm $U^{''}(M)$ is continuous with $M$ and 
vanishes exactly in the two-phase region.
 See Figs.\ \eqref{LS275} and\ \eqref{LS25} for an illustration.

The conclusion of this section is that, if the singularity of the threshold function  $\mathcal{\omega}$ at 
$\omega = \omega_0+$ is characterized by  the  exponent $\nu$ (cf  Eq.\ \eqref{simpl}) and if $d >2$, then
\begin{itemize}
 \item For $0 < \nu < \dfrac{4 - d}{2}$ the inverse compressibility $U''(M)$ is 
continuous at $M_0$, which constitutes a severe flaw of the theory.
 \item For $  \dfrac{4 - d}{2} \leq \nu$ the inverse compressibility $U''(M)$ is discontinuous at $M_0$
\end{itemize}

\section{\label{numerI} Numerical integration of the dimensionned RG flow equations}

\subsection{\label{algo} Algorithm}
In order to solve the EDP~\eqref{flowL} we must specify
\begin{itemize}
 \item{(i)} an arbitrary initial condition at $t=0$ ($k=\Lambda$)
\begin{equation}
 L_{\Lambda}(M) = L_i(M) \; \; \;  \forall M \in (-M_{\text{max}}, M_{\text{max}}) \; ,
\end{equation}
 where $M_{\text{max}} $ is some maximum magnetization.
 \item{(ii)} boundary conditions 
\begin{equation}
 L_{k}(\pm M_{\text{max}}) = L_b(k) \; \; \;   \forall k \in (k_{\text{min}}, \Lambda) \; .
\end{equation}
 where $k_{\text{min}} $ is some minimum value of the scale $k$.
 \item{(iii)} The functions $L_i(M)$ and $ L_b(k)$  \textit{a priori} arbitrary must be such that \cite{Ames} 
\begin{equation}
\label{compati}
 L_{i}(\pm M_{\text{max}}) = L_b(\Lambda)  .
\end{equation}
\end{itemize}
Under these 3 conditions the RG flow equation for $L_k$ have a unique solution. The choice of $L_i$ and $L_b$ relies
on physical grounds. At $k=\Lambda$ the cut-off function $R \propto \infty$ should diverge \cite{Wetterich,Delamotte}
such that $U_{\Lambda}
\equiv V_{\Lambda}$.  Thus, clearly,  $L_i(M)=\mathcal{L}(V^{''}_{\Lambda}(M)/\Lambda^2)$ must be imposed as an initial condition
even if $\Lambda$ is not large enough so that  $R_{\Lambda}$ is not actually infinite in the mathematical sense.

We propose to choose as boundary conditions the one-loop approximation for $ L_{k}(\pm M_{\text{max}})$. Most authors
 usually adopt free boundary conditions. At a large $M_{\text{max}}$ fluctuations are frozen and the one-loop 
approximation  should be pertinent. It is easy to see that at the one-loop level one has, up to an additional constant,   independent of the field  \cite{Zinn}
\begin{eqnarray}
 U_k^{\text{1-loop}}(M) & = & \frac{ \Gamma_k^{\text{1-loop}}(M)}{\Omega} \nonumber \\
&=& V_k(M) \nonumber \\
&+&  \frac{1}{2} \int_q \ln [ \frac{q^2 + V^{''}_{\Lambda}(M) + \widetilde{R}_k(q^2)}{\Lambda^2}]  \; ,
\end{eqnarray}
from which it follows that :
\begin{eqnarray}
 \partial_k U_k^{\text{1-loop}}(M) &=&  \frac{1}{2} \int_q  \frac{\partial_k \widetilde{R}_k(q^2)}{q^2 + V^{''}_{\Lambda}(M) + \widetilde{R}_k(q^2)} \nonumber \\
&=& 2  v_d k^{d-1} \mathcal{L}(V_k^{''}(M)/k^2) \; .
\end{eqnarray}
Therefore the one-loop approximation for $L_k(M)$ reads as
\begin{equation}
 L_k^{\text{1-loop}}(M) \equiv  \mathcal{L}(V_k^{''}(M)/k^2) \; .
\end{equation}
 
To summarize we impose
\begin{itemize}
 \item{(i)} initial condition :
$ L_i(M) =   \mathcal{L}(V_{\Lambda}^{''}(M)/\Lambda^2)    \text{ for all }  -M_{\text{max}}\leq  M \leq M_{\text{max}}$.
\item{(ii)} boundary conditions $ L_b(k) =   \mathcal{L}(V_k^{''}(\pm M_{\text{max}})/k^2)  \text{ for all }  0\leq  k \leq \Lambda$.
\end{itemize}
and we stress that the compatibility condition~\eqref{compati} is obviously satisfied hence
the RG flow equation can be solved safely at least  from a mathematical point of view.
\subsection{\label{num} Numerical experiments}

\begin{figure}[t!]
\includegraphics[angle=0,scale=0.55]{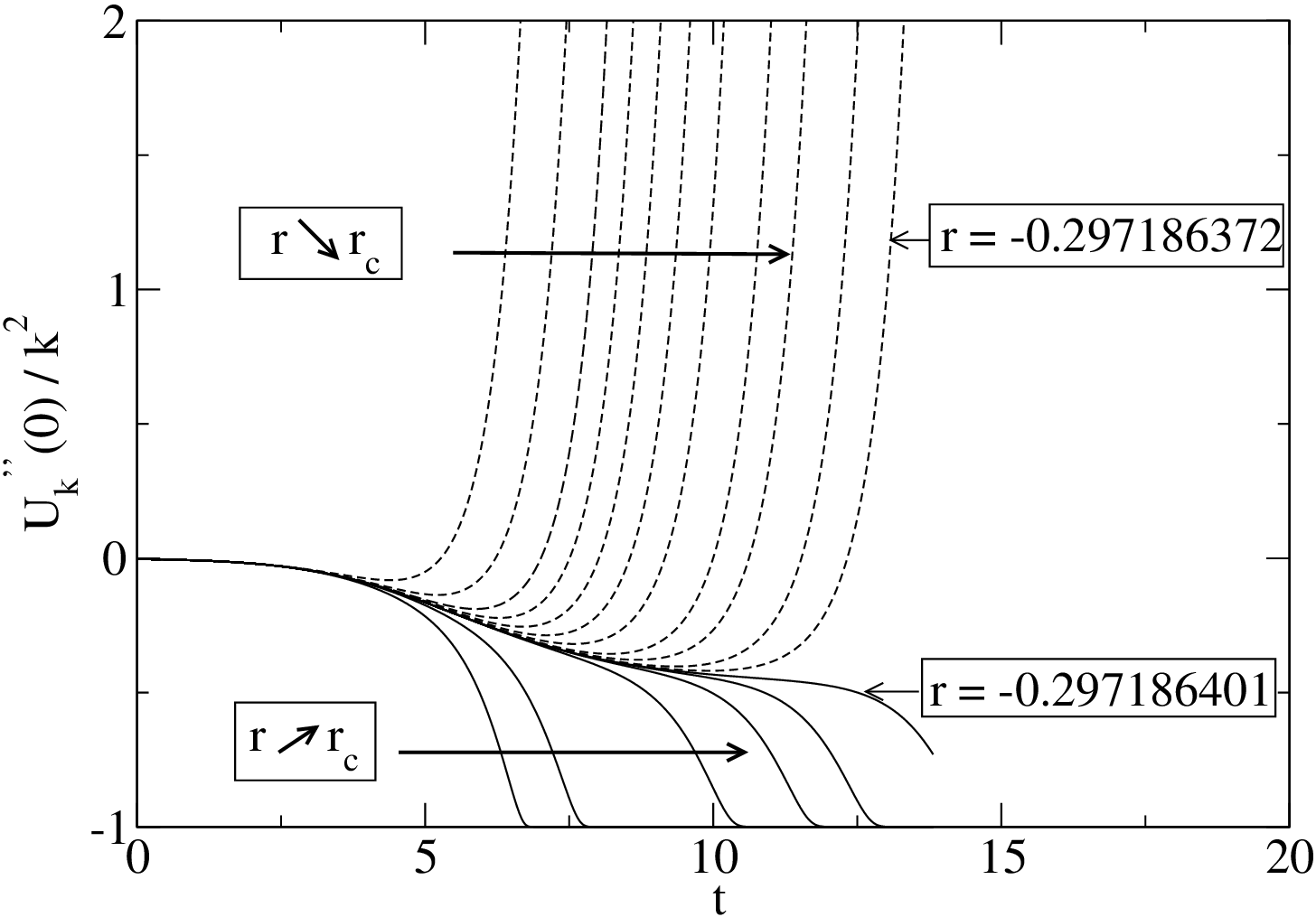}
\caption{
\label{figrc}  $U_k^{''}(M=0)/( k^2 \vert \omega_0 \vert ) $ as a function of the RG time ``t'' in the Sharp cut-off approximation.
$r_c$ is obtained by dichotomy. Dashed lines : $r > r_c$. Solid lines : $r < r_c$.
}
\end{figure}
The RG flow equation~\eqref{flowL}, completed with the initial and boundary conditions of Section~\ref{algo}, is a standard 
well-conditionned quasi-linear parabolic
EDP which can easily be solved numerically.
We   adopted  the
fully implicit predictor-corrector integration scheme of Douglas and Jones \cite{AmesI} which proved to be very efficient and precise.
Douglas and Jones have proved that their procedure leads to  an unconditional convergence
to the solution  and that  the truncation error is of order $\mathcal{O}(\delta t^2 + \delta M^2)$
where $\delta t$ is the discrete time-step and $\delta M$  the grid-step for the field.

We limited ourselves to the  potential  $V_{\Lambda}\left(\varphi\right)=\frac{1}{2!} r \varphi ^{2} +
\frac{1}{4!}g \varphi ^{4}$  with $\Lambda=10$ (arbitrary units) and $g=1.2$ being kept fixed  while varying $r$  in order to make 
some contact with 
previous numerical studies \cite{Bonanno}.   This is a well acceptable choice of initial condition provided one does not want to describe nonuniversal characteristics of, e.g., the Ising model.
The model was studied with the sharp, Litim and smooth cut-off regulators introduced
in Section~\ref{Thres}. Some studies with the simplified smooth cut-off described by Eq.~\eqref{simpl} with $\nu=1/2$ were also performed
to compute $\chi(M)$ in the vicinity of $ \pm M_0$.
Since no numerical simulations are available for this continuous version of the $\varphi^4$ model we tried
to extract from our data similarities as well as differences between the various versions of the LPA. 

Typically we retained $M_{\text{max}}=8$, after checking  that the results
were not influenced by the  boundaries conditions. The time-step  was typically $\delta t=2. 10^{-4}$ and the field-step ranging from
 $\delta M=2. 10^{-3}$ to  $\delta M=2. 10^{-6}$ according to the problem at hand. In all cases the integration of the EDP was stopped
at some $k_{\text{min}}$ between $ \sim 10^{-6}$ and   $ \sim 10^{-10}$ after checking  that the solution did not evolve any more. 
Typically about $N_t \sim 10^5$ time steps were necessary to
reach convergence. For the smooth cut-off it was moreover
necessary to fit the function
$\omega(M)$ defined as the inverse of function $\mathcal{L}$ (cf Eq.~\eqref{L}), its derivatives and inverse yielding 
unavoidable systematic errors and a  reduction of the precision on the data.


\subsubsection{\label{num1} Thermodynamic potentials below $r_c$}
We display in Fig.~\ref{lnL} the evolution of  threshold function $L_k(M)$ when the scale-k of the RG flow 
decreases  from $\Lambda=10$
 to $k_{\text{min}}=10^{-4}$. The curves have been obtained with Litim's regulator; other approximations give 
similar curves, except the 
Sharp cut-off since $L_k $ can become negative in this case. 
The universal behavior~\eqref{universal} of the LPA is exemplified in  Fig.~\ref{Luniv}
in the appropriate reduced  variables.

 The behavior of the potential $U_k(M)$ which can be extracted from  $L_k(M)$ is displayed in Fig.~\ref{figU}.
Note that the second derivative  the true Gibbs potential $\overline{U}''_{k}(M)$ of the k-system remains positive as
we shown in sec~\ref{LPA} (cf. Eq.~\eqref{Ubar}).  For the considered state and at the lowest $k= k_{\text{min}}=10^{-4} $
 where we stopped the
integration of the RG flow one finds 
$M_0=0.9356$ and the inverse susceptibility $\overline{U}''_{k_{\text{min}}}(M)$
jumps from $0.3142$ at $M=M_0 + \delta M $ to  $0.4512 \; 10^{-10}$  at  $M=M_0 -\delta M$.
 Lowering the value of $k_{\text{min}}$ doe not change the value of the spontaneous magnetization
$M_0$ (for a given grid step) while  $\overline{U}''_{k_{\text{min}}}(M)$ may be given any arbitrary small value
for $\arrowvert M \arrowvert<M_0$.

The approach to convexity is also well illustrated by the anti-clockwise rotation of
the plateau  of $U_k^{'}(M)$, \textit{i.e.} the precursor of the magnetic field,  towards
an horizontal segment as $k \to 0$ in agreement with Eq.~\eqref{U}. Finally one notes that
all the  $U_k(M)$
behaves as parabola inside the coexistence curve, according to Eq.~\eqref{universal},  as soon as
$k$ is sufficiently small.


\subsubsection{\label{num2} Continuity/discontinuity of the susceptibility}
\begin{figure}[t!]
\includegraphics[angle=0,scale=0.55]{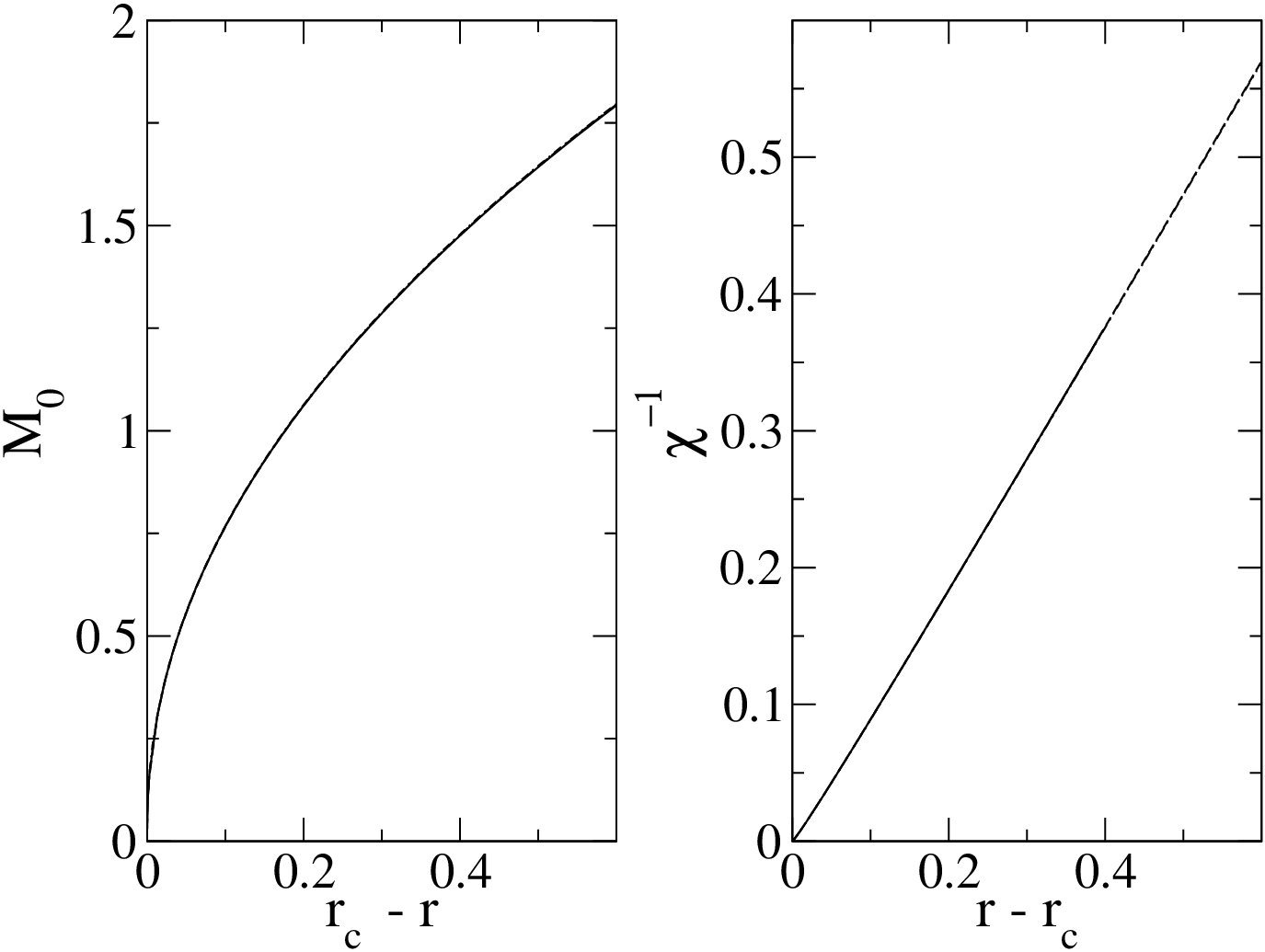}
\caption{
\label{figcompa} 
Left : magnetization versus $r_c-r$ for Litim (solid line), sharp (dashed line), smooth cut-off (dashed-dotted line) regulators.
Right : inverse susceptibility $\chi^{-1}= U^{''}_{k=0}(M=0)$  versus $r - r_c$, same symbols. In both cases 
$\varphi^4$ model with  $\Lambda=10$ with $g=1.2$ 
}
\end{figure}

Here we check numerically the conclusions of Sec~\ref{LPA} concerning the continuity or discontinuity of the susceptibility 
at $M=  M_0$ in the various versions of the LPA.
Fig~\ref{LS} displays the inverse susceptibility $U^{''}_{k=0}(M)$ in the vicinity of $M_0$ for the sharp, Litim's, and exponential smooth cut-off    
in $d=3$ dimensions. A relatively small field-step $\delta M=2. 10^{-6}$
is required to be convinced of the (dis)continuity of the curves. Although such numerical checks 
cannot be retained as mathematical proofs
\textit{stricto sensu}, the theoretical predictions, \textit{i. e.} continuity of $U''_{k=0}(M)$ at $M=M_0$ 
for the sharp cut-off and discontinuity for the Litim and smooth exponential  regulator are quite well
supported by the numerical outcomes.

Figs~\ref{LS325},~\ref{LS3},~\ref{LS275} and~\ref{LS25} display the curves  $U''_{k=0}(M)$ obtained
numerically in the case of the  simplified smooth cut-off
threshold function, \textit{i.e.} 
$ \mathcal{L}(\omega) = K_0 \; (\omega - \omega_0)^{-1/2} $  for dimensions of space
$d=3.25$,  $d=3$,  $d=2.75$, and $d=2.5$ respectively.
Recall that the behavior of any theory involving a smooth regulator is expected to be of that type.
In all cases $\Lambda=10$, $g=1.2$ and
$M_{\textrm{max}}=8.$ while the  parameter $r$ of the $\varphi^4$ potential has been adjusted 
for a spontaneous magnetization 
$M_0 \sim 0.05$. 
For each dimension ''$d$'' we considered several values of field-step $\delta M$ ranging from  $\delta M=2. 10^{-3}$ to
 $\delta M=2. 10^{-6}$ 
corresponding resp. to a sampling of $N_M=4.10^3$ to $N_M= 4. 10^6$ points in the interval $(0,M_0)$
in order to emphasize the numerical difficulty to check this (dis)continuity.
As expected (see Sec~\ref{LPA} )
a spectacular change of behavior at $d=3$. From continuous at low dimensions  (cf the analysis of Sec~\ref{I}) 
$U''_{k=0}(M)$ becomes discontinuous for $d \geq 3$. Generally at most a single point seems to survive half the way
the two discontinuity points whatever the
tiny the value of $\delta M$ considered. In particular note that,
exactly at $d=3$,  $U''_{k=0}(M)$   ``looks'' indisputably discontinuous at $M=\pm M_0$
as anticipated by the analysis of sect~\ref{II}.


\subsubsection{\label{num3} Determination of the critical point and the critical exponents}


It follows from Eq.~\eqref{U} that the quantity $ y_k = U''_k(M=0)/(k^2 \vert \omega_0 \vert)$ is well suited to discriminate states
at a temperature above $r_c$ from those  at a temperature below $r_c$. 
Indeed, as $k\to 0$,  $y_k$ diverges to $+\infty$ at $r>r_c$ ( since the inverse compressibility $U''_k(M=0)$ 
is finite for $r>r_c$),
while it tends to $-1$ for subcritical states, ( cf Eq.~\eqref{U}).
A dicothomy procedure  then yields the precise determination of the critical temperature as exemplifies by Fig.\ref{figrc}
which displays results only  for the ultra-sharp cut-off since in the other variations of the LPA (Litim's or smooth cut-off) 
the bunch of curves is roughly similar and brings no new information.
The results for $r_c$ are reported in table~\ref{tablerc}. For a given approximation they depend slightly on the field-step $\delta M$
but as soon as $\delta M$  is as small as $2. 10^{-4}$ we noted no  effect on $r_c$, up to its $12th$ digit,  by decreasing its magnitude.
Note that we can ascertain   $ \sim 8$ exact  digits on the value of $r_c$ from  the data 
obtained  with $\delta M < 2. 10^{-4}$. 
\begin{table}[t!]
\caption{\label{tablerc} Influence of the field-step $\delta M$ on the critical  parameter  $r_c$ 
in all the variations of the LPA considered in this work : 
Sharp, Litim and exponential smooth cut-off.  $r_c$ was determined by dicothomy as discussed in Sec~\ref{num3}.
Reported data concern  the  $\varphi^4$ potential defined at $\Lambda=10$,  $g=1.2$  and $r=r_c$. The RG  flow equations have been interrupted
at $k_{\text{min}}=10^{-10} $.
In all cases the time-step is $\delta t = 2. 10^{-4}$. We report 10 stable digits, numerical uncertainties only affect the 11th.}
\begin{tabular}{|l|c|c|c|} 
\hline
    $\delta M $      &   $r_c$ (Litim)           &   $r_c$ (Sharp)        &   $r_c$ (Smooth)               \\ \hline
$ 2. 10^{-3}$        &  -0.1969317598       &   -0.2971859571      &    -0.4120948764               \\ \hline
$ 2. 10^{-4}$        &  -0.1969322088       &   -0.2971863642      &    -0.4120953249               \\ \hline
$ 2. 10^{-5}$        &  -0.1969322133       &   -0.2971863684      &    -0.4120953292               \\ \hline
\end{tabular}
\end{table}
\begin{table}[h!]
\caption{\label{tab_ind} Critical exponents of the magnetization and the susceptibility, resp. $\beta$
and $\gamma$, obtained by the numerical experiments of Sec~\ref{num3}. 
The number(s)  in brackets denote the linear
regression error on the last digit(s). $|r -r_c|$ was varied from $10^{-7}$ to $10^{-4}$.
The critical  exponent $\alpha$ was obtained 
by using Rushbrook's equality $\alpha + 2 \beta + \gamma =2$.}
\begin{tabular}{|l|l|l|l|} 
\hline
                                   &       Litim                    &      Sharp                   &   Smooth                   \\ \hline
$   \beta      $               &       0.3327(8)           &      0.3532(9)             &    0.3352(8)               \\ \hline
$   \gamma $               &       1.2768(14)          &     1.3292(26)           &    1.2781(14)             \\ \hline
$   \alpha    $               &       0.0578(32)          &    -0.0356(44)           &   0.0515(32)              \\ \hline
\end{tabular}
\end{table}
Quite noticeable  is the dispersion of the values of $r_c$ corresponding to different cut-off functions  which casts some
doubts to the validity of the LPA to predict quantitative non universal results. This is an important 
detail that the non-universal parameters appear to depend strongly on the initial value considered at $k=\Lambda$. 
If one changes something (a cutoff function for example), then the nonuniversal characteristics of the system changes. Moreover 
the MF approximation injected as an initial condition for the flow is a too crude approximation.
From this respect our results are only but superficially  at odds with those obtained recently  by Dupuis and Machado for the Ising and
lattice $\varphi^4$ models  where
an excellent agreement (within a few percents) between the LPA prediction for  $r_c$ and the MC data was found \cite{Dupuis}.
However in their work all these authors adopt a slightly modified  version of the NPRG-LPA in which the effective action at
 $ k= \Lambda$ is not taken as the mean field result but as the exact one (at this scale)
and where integrations over the continuous momenta $q$
is replaced by a summation over the vectors $\vec{q}$ of the  first Brillouin zone.
Similarly in the domain  of the theory of liquids, the works reported by Reatto \textit{et al.} attest good agreement between the LPA
and the MC data \cite{ReattoI,ReattoII,Reatto0}; in this case the exact physics of a reference system (the hard spheres fluid) is 
injected in the theory. 

\begin{figure}[t!]
\includegraphics[angle=0,scale=0.55]{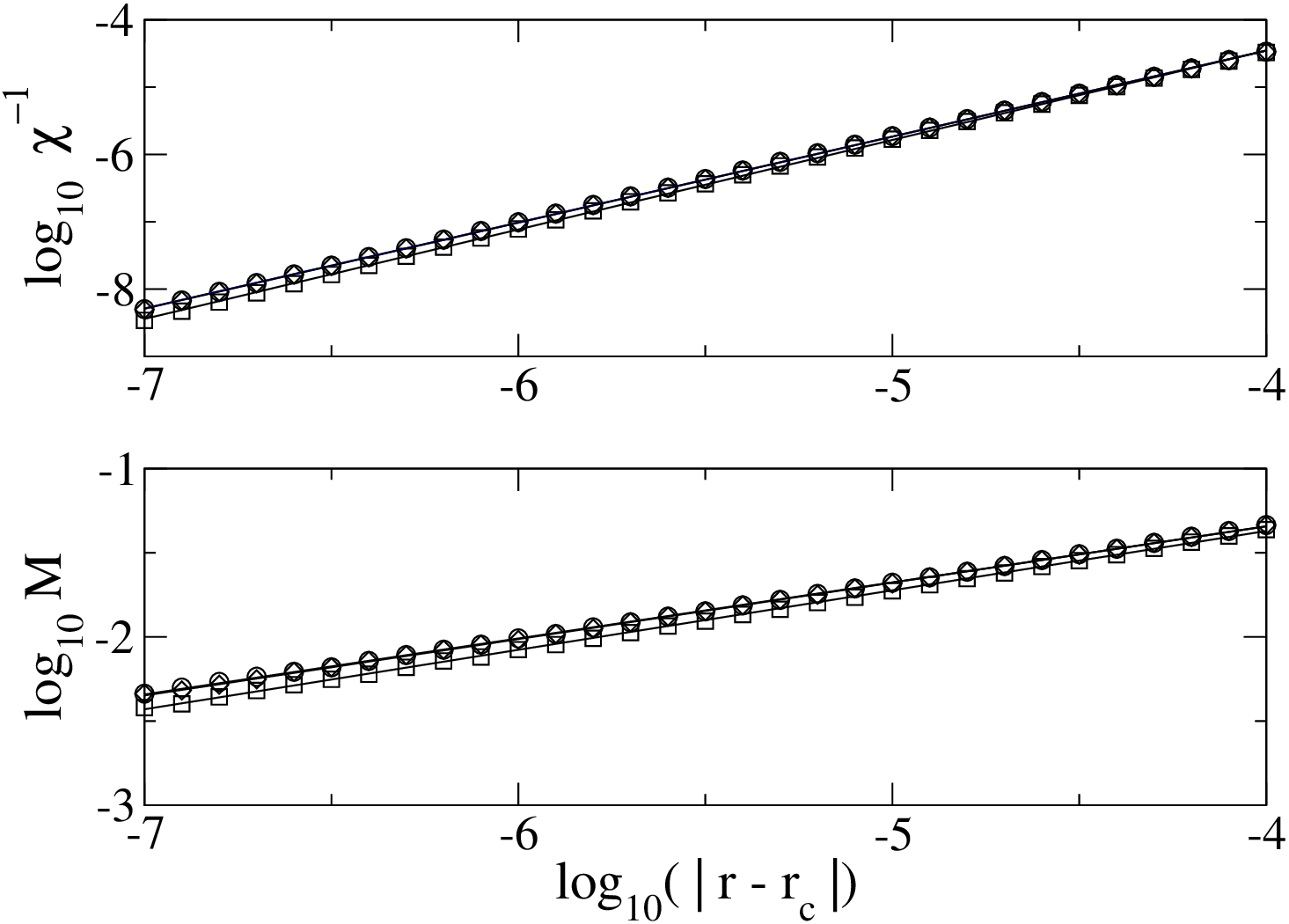}
\caption{
\label{index} 
Bottom : logarithm of the magnetization versus $\log_{10} (\arrowvert r_c-r \arrowvert )
$ for Litim (circles), sharp (squares), 
exponential smooth cut-off (diamonds) regulators. The lines are linear regressions of the data.
Top : logarithm of the inverse susceptibility $\chi^{-1}= U^{''}_{k=0}(M=0)$  versus
$ \log_{10} (\arrowvert r_c-r \arrowvert ) $, same symbols. In both cases 
$\varphi^4$ model IN  $d=3$ with $\Lambda=10$,  $g=1.2$, and $M_{\textrm{Max}}=8$.
}
\end{figure}

However when either the spontaneous magnetization $M_0(r)$  (for $r <r_c$)  or the inverse susceptibility $U''(M=0)$ (for  $r >r_c$)
 are reported on a graph as  functions of $r -r_c$ rather than versus $r$
all the curves obtained by means of different regulators collapse on a single, approximatively universal  one, at least at large scale. 
This striking observation is exemplified in Fig.~\ref{figcompa}. In first approximation the effect of the cut-off seems to be a simple shift on $r_c$.
If we abandon the Sirius point of view and zoom in the vicinity of $r\sim r_c$
the various routes yield in fact different behaviors since the critical exponents differ slightly. A series of numerical experiments
which are resumed in Fig.~\ref{index} allows a rough determination  of the critical exponents of the magnetization -$\beta$-
and  the susceptibility -$\gamma$-. As apparent in Table~ III the values obtained for these two exponents in the case of the
Litim and Smooth cut-off regulators
are in good agreement but they differ quite significantly from those obtained for the sharp cut-off. By passing we note that  Rushbrook's equality 
yields a negative $\alpha$
(exponent of the specific heat) in the case of the sharp cut-of, a well-known flaw of this approximation. 
A more stringent discussion
will be given in next section where the exponents will be calculated  with a high precision.

\section{\label{adim}Integration of the adimensionned RG equations}

.

\subsection{\label{adim_eq} Adimensionned RG flow equations}
In order to study fixed point solutions and the spectrum of critical exponents of the LPA
we need to work with  dimensionless  potentials and fields. Since $[U]=k^d$ and $[M]=k^{-1 + d/2}$
we define the dimensionless magnetization   $\widetilde{M}=k^{1- d/2}\, M$
and the dimensionless
potential $\widetilde{U}(\widetilde{M})= k^{-d}\,U(M)$, so that the RG flow Eq.~\eqref{flowU2} takes now the form 
\begin{equation}
 \label{flow_u_adim_0}
\partial_{t}\widetilde{U} = d\, \widetilde{U}+ (1-\frac{d}{2})\, \widetilde{M} \, \widetilde{U}'
-2 v_d\, \mathcal{L}(\widetilde{U}'') \; ,
\end{equation}
 in which the ' designates a derivative with respect to $x$.
Further simplification can be obtained by introducing finally reduced variables.
In the general case  one defines
$x=\widetilde{M}/\sqrt{2 v_d}$ and $u = \widetilde{U}/(2\, v_d)$. Litim case will be our exception
with the choice $x=\widetilde{M}/\sqrt{4 v_d/d}$, $u = d\widetilde{U}/(4\, v_d)$ and the redefinition
valid henceforth $\mathcal{L}(\omega) = 1/(1 + \omega)$. With these notations 
we have now in any case :
\begin{equation}
 \label{flow_u_adim}
\partial_{t} u = d \, u + (1-\frac{d}{2}) \,x \, u' - \mathcal{L}(u'') \; . 
\end{equation}

Similarly one introduces  the reduced  threshold potential   $ l(x,t)\triangleq L_k(M) =\mathcal{L}(u''(x,t)) $
 which obeys the following PDE
\begin{equation}
 \label{flow_l_adim}
\omega'(l) \, \partial_{t} l = 2 \omega(l)  + (1-\frac{d}{2}) \, x \, \omega'(l) \, l' - l'' \; ,
\end{equation}
where we recall that $\omega = \mathcal{L}^{-1}$ is the inverse of function 
$\omega \rightarrow \mathcal{L}(\omega)$. Eq.~\eqref{flow_l_adim}
was obtained by differentiating twice Eq.~\eqref{flow_u_adim} with respect to the
field variable $x$.

\subsection{\label{fixed_point} Fixed points and exponents}
Fixed point solutions $u^{\star}(x)$ and $l^{\star}(x)$ of Eqs.~\eqref{flow_u_adim}
and~\eqref{flow_l_adim}   resp.  are peculiar solutions of the ordinary differential equations (ODE)
 \begin{subequations}
\begin{eqnarray}
 0 &=& d u^{\star} +(1-\frac{d}{2}) \, x \,  u^{\star\, '} -\mathcal{L}(u^{\star\, ''})  \; ,  \label{fixu}\\
 0 &=& 2 \, \omega ( l^{\star})  +(1-\frac{d}{2}) \, x \, \omega^{'} (l^{\star}) \, l^{\star\, '} -
l^{\star \,''} \; ,\label{fixl}
\end{eqnarray}
\end{subequations}
respectively. Note that if  $u^{\star}$ is a solution of Eq.~\eqref{fixu} then $l^{\star}(x)=\mathcal{L} \circ u^{\star\, ''}(x) $ is
a solution of Eq.~\eqref{fixl}; similarly from a solution  $l^{\star}$ of~\eqref{fixl} one builds 
$u^{\star\, ''}(x)= \omega \circ l^{\star}(x) $ to get a solution of ~\eqref{fixu}. 
Because of  the  assumed  $\mathbb{Z}2$ symmetry  we will consider only even solutions $u^{\star}(x)=
u^{\star}(-x)$ of  Eq.~\eqref{fixu}. Then, since  $l^{\star}(x)=\mathcal{L} \circ u^{\star \, ''}(x)$, 
$l^{\star}(x)$ is also an even solution of Eq.~\eqref{fixl}.

Once an even fixed point has been found (analytically or numerically) one investigates the behavior of the solutions of
Eqs.~\eqref{flow_u_adim} and~\eqref{flow_l_adim} in the vicinity of this fixed point. As usual, we write
$u (x,t) = u^{\star} (x) + h_{\lambda} (x) \exp(\lambda t)$ and consider $h_{\lambda}(x)$ as small. Linearizing Eq.~\eqref{flow_u_adim}
with respect to $h_{\lambda}$ leads to the eigenvalue equation :
\begin{subequations}
\begin{equation}
\label{vp_u}
0=  (d-\lambda) \, h_{\lambda} + (1 - \frac{d}{2}) \, x \, h_{\lambda}^{'} -\mathcal{L}^{'}(u^{\star\, ''}) h_{\lambda}^{''} \; .
\end{equation}

Similarly writing  $l (x,t) = l^{\star} (x) + g_{\lambda} (x) \exp(\lambda t)$ and linearizing the full RG flow equation with respect
to $g$ yields.
\begin{widetext}
\begin{equation}
\label{vp_l}
0 =  g_{\lambda} \, \omega^{'}(l^{\star}) \, (2 - \lambda)  + (1 - \frac{d}{2}) \, x \left\{ \omega^{'}(l^{\star}) \, g_{\lambda}^{'} +
\omega^{''}(l^{\star}) \,  l^{\star \, '} \, g_{\lambda} \right\} - g_{\lambda}^{''}
\end{equation}
\end{widetext}
\end{subequations}
A first glance, since 
$$l(x,t) = \mathcal{L}(u^{''}) \sim  \mathcal{L}(u^{\star ''})+
 \mathcal{L}^{'}(u^{\star ''})h_{\lambda}^{''}(x) \exp(\lambda t)$$ the two spectra are identical
and the eigenfunctions  related by 
\begin{equation}
\label{ro}
 g_{\lambda}(x)=\mathcal{L}^{'} \circ u^{\star \, ''}(x) \, h_{\lambda}^{''}(x) \, .
\end{equation}
Strictly speaking this  is true only if $h_{\lambda}^{''}(x)$ is not identically equal to 0. Thus if it
turns out that $h_{\lambda}(x)$ is either  a constant or a linear function of $x$ then  eigenvalue $\lambda$
does not belong to the spectrum of the RG operator acting on $l$. As well known, odd and even eigenvectors 
form two mutually orthogonal linear subsets  and will be both considered in the sequel.

\subsection{\label{trivial} Trivial solutions}
Such trivial solutions exist whatever the type of fixed point; they are more easily detected on the eigenvalue
problem~\eqref{vp_u}, \textit{i.e.} that attached to the linearization of the RG about  $u^{\star}(x)$.
Let us rewrite Eq.~\eqref{fixu} for $u^{\star}$
as well as the equation for its derivative $f^{\star}=u^{\star \, '} $; one has
 \begin{subequations}
\begin{eqnarray}
  0 &=& d u^{\star} +(1-\frac{d}{2}) \, x \,  u^{\star\, '} -\mathcal{L}( u^{\star\, ''})  \, , \\
  0 &=& (1 + \frac{d}{2}) f^{\star} + (1 - \frac{d}{2}) \, x \,  f^{\star \, '} -\mathcal{L}^{'}( u^{\star\, ''}) 
\end{eqnarray}
\end{subequations}
Comparing these equations to the eigenvalue problem~\eqref{vp_u} one readily sees that
\begin{itemize}
 \item[(i)] The constant  $h_{\lambda}(x)= h_0$ is a trivial even eigenfunction of~\eqref{vp_u} with the 
               eigenvalue $\lambda = d$ 
 \item[(ii)] $h_{\lambda}= f^{\star} (x)$ is also an  eigenfunction of~\eqref{vp_u}  associated to the  eigenvalue
                $\lambda =-1+ d/2$. Since $ u^{\star}(x)$ should be even then  $ f^{\star}(x)$ is an odd eigenvector;
                 according to  Wegner it is associated with  the shift operator \cite{Wegnerb}.
 \item[(iii)] $h_{\lambda}= x $ with  $\lambda =1+ d/2$ is another trivial odd eigenvector associated with the magnetic field.
\end{itemize}
From the remark of the previous section it follows that only the trivial odd eigenvalue 
$\lambda =-1+ d/2$ survives in the eigenvalue problem~\eqref{vp_u}, \textit{i.e.}
that attached to the linearization of the RG flow about  $l^{\star}(x)$.
According to~\eqref{ro} the corresponding eigenvector is the odd function
$g_{\lambda=-1 +d/2}(x)=\mathcal{L}^{'} \circ f^{\star \, '}(x) \, f^{\star \, ''}(x)$.

\subsection{\label{Gauss} Gaussian fixed point}
We show now that, provided that $d>2$,  the LPA admits a Gaussian fixed point,  with the usual spectrum
of  exponents, irrespective to the type of cut-off.
The fixed-point equation for $u^{\star}$, \textit{i.e.} Eq.~\eqref{fixu},  admits $u^{\star \, ''}(x)=0$ as a special solution.
 By integration it gives
 $u^{\star \, '}(x)=0$  ($\mathbb{Z} \, 2$ symmetry) and $u^{\star }(x)=u_G$ with $d\, u_G= \mathcal{L}(0)$; where
 $\mathcal{L}(0)=0, \, 1$ for the sharp and Litim's cut-off respectively  and,  numerically, $\mathcal{L}(0)=5.9973827$ for 
our smooth cut-off.
Obviously  $l^{\star}(x)=l_G$ is the related special solution of Eq.~\eqref{fixl} provided $l_G=d\, u_G$.

The linearized eigenvalue problem for $u(x,t)$ reads 
\begin{equation}\label{sto}
 (d-\lambda) \, h(x) + (1-\frac{d}{2}) \, x \, h^{'}(x) - \mathcal{L}^{'}(0)\, h^{''}(x)=0 \;,
\end{equation}
where $ \mathcal{L}^{'}(0)<0$ (cf  $\mathcal{L}^{'}(0)= - 1$ for Sharp and Litim's cut-off, while, numerically
 $\mathcal{L}^{'}(0)= - 1.37960752$ for the smooth-cut-off). The change of variables
 \begin{subequations}
  \begin{align}
   h(x) &= H(y)           &          y &=\beta x \\
 \beta & = \sqrt{\dfrac{4}{(2-d) \, \mathcal{L}^{'}(0)}}  &  2\, n & = \dfrac{4(d-\lambda) }{d-2}
  \end{align}
 \end{subequations}
allows to rewrite Eq.~\eqref{sto} as Hermite equation :
\begin{equation}\label{H}
 H^{''}(y) - 2y \, H^{'}(y) +2n \, H(y) =0 \; ,
\end{equation}
If we request the potential to be bounded by polynomials,  $n$ must be restricted to an integer.
Then Eq.~\eqref{H} becomes the equation defining Hermite Polynomials
$H_n(y)$. The parity of $H_n(y)$ being that of $n$, the spectrum of even eigenvalues is given by
\begin{equation}
 \lambda_p = d -p\, (d -2) \; ,
\end{equation}
with $n = 2\, p$ and $p=0, 1, 2 \ldots$.
The trivial relevant operator $H_0$ with $\lambda_0=d$ is indeed present in the spectrum as discussed in sec~\eqref{trivial}.
Whatever the type of cut-off we recover the well-known result :
in $d=4$ we find one relevant operator ($H_2$ with $\lambda_1=2$), a marginal operator ($H_4$ with $\lambda_2=0$)
and the first irrelevant operator with $\lambda_2=-2$ ($H_6$) while in $d=3$  there are two relevant operators 
($H_2$ and $H_4$
with $\lambda_1=2$ and $\lambda_2=1$, respectively) and a marginal operator ($H_6$ with  $\lambda_3=0$), 
while the first irrelevant operator
is $H_8$ with  $\lambda_4=-1$.
Some remarks are in order.
\begin{itemize}
 \item An analysis similar to that of ref.\cite{Hasen} reveals that the marginal operators become irrelevant at the quadratic order.
\item The trivial eigenvalue $\lambda_{p=0}= d $ disappears from the spectrum of Eq.~\eqref{vp_l}. The eigenvectors of Eq.~\eqref{vp_l}
are given by $g_{\lambda_p}(x)=(4/(2-d))\, H_{2p}^{''}(y)$ as follows from Eq.~\eqref{ro}.
\item The odd spectrum is given by $\lambda_p=d -(p +1/2)(d-2)$, $p$ an integer. It includes the trivial  solution
 $\lambda_0=1+d/2$. The other trivial odd  eigenvalue  $ \lambda=-1+d/2$ is absent accidently from the spectrum, since
 it should correspond to  a zero eigenvector ($h_{\lambda} = f_G^* \equiv 0$).
\end{itemize}

\subsection{\label{non_trivial} Non Gaussian fixed point}

In this section we focus on the non Gaussian fixed point in $d>2$.
Recently, the LPA fixed point equation for $u^{\star}(x)$ have been solved with a very high numerical precision
for the sharp and Litim's cut-off  \cite{BerGiacoI,BerGiacoII}.
Here we report numerical solutions  for $l^{\star}(x)$ only in dimension $d=3$ ;  the three cut-offs
considered in  this paper were  examined and compared. 
Eq~\eqref{fixl} can be solved by the shooting method with boundary conditions 
imposed either at the origin  $x=0$ 
or at $x=\infty$ \cite{BerGiacoI,BerGiacoII}, hence the names given to the two methods considered below.

\subsubsection{\label{ab} \textit{ab origine}}

\begin{figure}[t!]
\includegraphics[angle=0,scale=0.35]{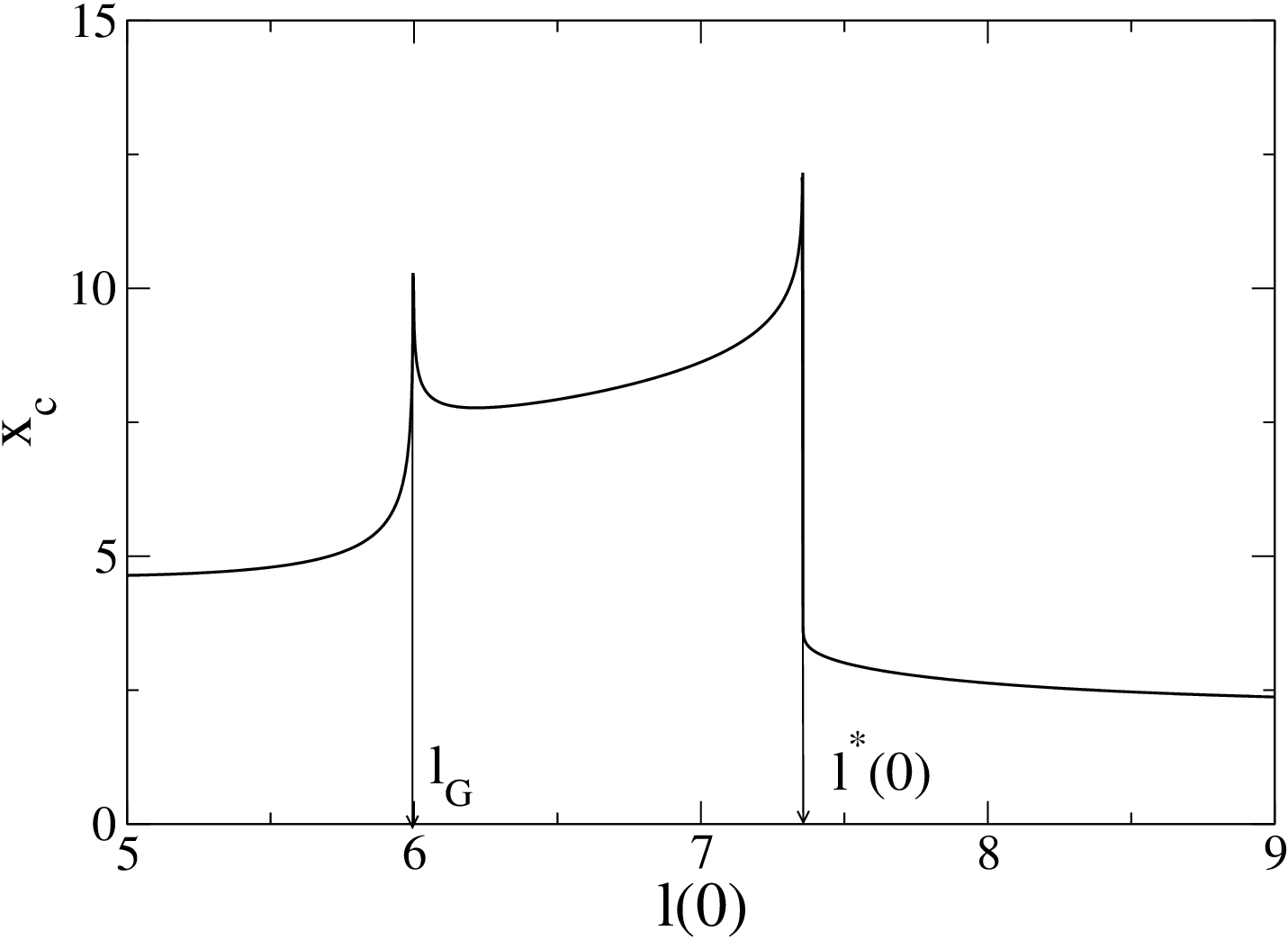}
\caption{
\label{Phi_L} 
Singular field $x_c$ as a function of the initial value $l(0)$ of 
Eq.~\eqref{fixl} in the case of the Smooth cut-off.
}
\end{figure}

\begin{table}[h!]
\caption{\label{tab_ab} Results from the shooting method ``ab origine''.}
\begin{tabular}{|l|r|r|r|} 
\hline
                                    &       Litim                            &      Sharp                       &   Smooth                   \\ \hline
$  l^{\star}(0)         $     &   0.122859820243702      &   0.61903040294652    &     7.355923051         \\ \hline
$  u^{\star \, ''}(0)  $     &    -0.186064249470314    &    -.46153372011621    &     -.7995141985        \\ \hline
$   x_c                  $      &  20.644305503116           &    94.128646935418     &  22.95208767             \\ \hline
\end{tabular}
\end{table}

The LPA fixed point equation~\eqref{fixl} is a second order ODE for the function $l^{\star}(x)$.
Here we solve it numerically in $d=3$ by providing two initial conditions at $x=0$. The first one, $l^{\star \, '}(x)=0$ ensures
the parity $l^{\star}(-x)=l^{\star}(x)$  required by $Z_2$ symmetry.  
It is now well-known that an arbitrary value of $l^{\star}(0)$ yields a solution singular at some finite value $x_c$ of the field.
At $x_c$ we have  $l^{\star}(x_c)=0$ for the Smooth or Litim's cut-off and $l^{\star}(x_c)=-\infty$ for the sharp cut-off. These singular values
for $l^{\star}(x_c)$ correspond to $u^{\star \, ''}(x_c)= + \infty$. Actually, the general solution of~\eqref{fixl} involves
a moving singularity of the form 
 \begin{subequations}
  \begin{align}
   l(x) & \sim K_1 \sqrt{d-2} \, (x_c-x)^{1/2} \, x_c^{1/2} & \textrm{Smooth, Litim} \\
 l(x)   & \sim \ln \left((x_c -x) \, x_c \, (d/2-1) \right)                               &  \textrm{Sharp}
  \end{align}
 \end{subequations}
where, in the case of Litim's cut-off  $K_1=1$ (the behavior near the singularity $x_c$ is driven by the asymptotics
at infinity of the function $\mathcal{L}(u^{''})$, cf Eq.~\eqref{asympto2}).
Figure~\ref{Phi_L} displays the variation of the singular point $x_c$ with the initial condition $l(0)$ in the case of the Smooth cut-off 
(similar curves are obtained for the Sharp and Litim's cut-off). Two peaks where
$x_c$ diverges to $\infty$ can be noticed. The one on the left corresponds  the Gaussian
fixed point  where $l(x)$  is a constant with $l(0) = l_G=\mathcal{L}(0)=5.9973827$ (see section~\ref{Gauss}).
The one one the right corresponds to the Wilson-Fisher fixed point with  $l(0) =  7.355923051$ that we are looking at..

Our requirement is that the physical solution must be non singular on the entire range $x \in (0, \infty)$ so
we must push $x_c$ to infinity by adjusting the value of $l(0)$ by a dichotomy process in the vicinity of the right
peak of figure~\ref{Phi_L} \cite{Felder}.
Of course high precision ODE solvers are required for this kind of study.

We solved equation~\eqref{fixl} as well as all the ODE of this paper with the DOPRI853 algorithm of Hairer \textit{et al.} \cite{Hairer}
 which is an explicit Runge-Kutta integrator of order ``8'' and order ``5'' embedded, with adaptive step-size. 
We imposed  relative and absolute 
errors of $10.^{-18}$ and $10.^{-25}$ respectively; the code was written in FORTRAN90 in quadruple precision.
Even with this high technology it is impossible to obtain very large values of $x_c$ by tuning $l(0)$. Our 
results are summarized in Table~\ref{tab_ab}. Our results for $u^{\star \, ''}(0)$ deduced of our result for 
 $l^{\star}(0)$ agree within 14 figures with  those of ref.\cite{BerGiacoI,BerGiacoII} in the case of Litim's and sharp cut-off
regulators.
The data reported in Table~\ref{tab_ab} for the smooth cut-off are much less precise due to the use of fits for computing
 function $\mathcal{L}(\omega)$. The complicated behavior of function $l^{\star}(x)$ is exemplified in fig.~\eqref{l_fix} (red curve);
We displayed only the result for the smooth cut-off, Litim's case is similar while for the sharp cut-off $l^{\star}(x)$
 tends to $- \infty$ at $x_c$ instead of ``0''.

This \textit{ab origine} method is rather deceptive but however usefull to check the data obtained by the 
 \textit{ad originem} shooting method that we discuss now. 
\subsubsection{\label{ad} \textit{ad originem}}
\begin{figure}[t!]
\includegraphics[angle=0,scale=0.45]{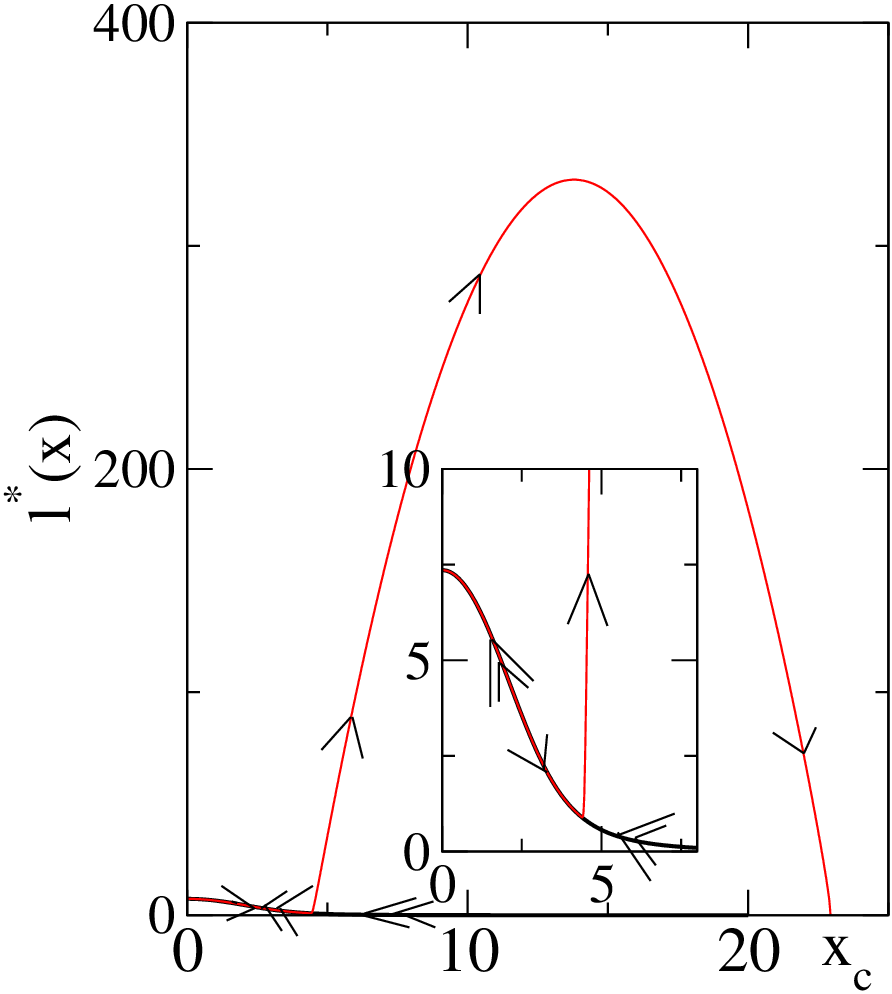}
\caption{
\label{l_fix} 
Fixed point solution $l^{(\star)}(x)$ for the smooth cut-off.
Red line : shooting \textit{ab origine}, black line : shooting \textit{ad originem}.
}
\end{figure}

The existence of a moving singularity at $x_c$ which seems impossible to ``push`` at infinity suggests to
solve Eq.~\eqref{fixl} with initial conditions at infinity, at least $x_{\textrm{max}}$ large, towards $x=0$. The analysis of
the asymptotic solutions of ~\eqref{fixl} for $x \to \infty$ shows the existence of power law solutions. This second family of solutions
with regular scaling properties must obviously be preferred to the singular solutions discussed in  section~\eqref{ab}.
One can show that, asymptotically, for $x \to \infty$ one has, for the smooth cut-off regulator
\begin{subequations}
 \begin{align}
\label{linf}
  l^{\star}(x)  &=  b_l \, x^{-\beta_l} + \frac{C_2}{C_1}b_l ^2 \,x^{-2 \beta_l}  +\nonumber \\
                   & +\left(  \frac{C_3}{C_1} + \frac{C_2^2}{C_1^2}\right) \, b_l ^3 \,x^{-3 \beta_l}
                 +\mathcal{O}(x^{-4 \beta_l}) \; ,
 \end{align}
with $\beta_l = 4/(d-2)$ and $b_l$ an arbitrary coefficient. Recall that $C1, \, C2, \,  C3$ enters the asymptotic behavior~\eqref{asympto3}
of $\omega(l^{\star})$ as $l^{\star}  \to 0$ and are given in table~\ref{constantes}.
 Litim's case can be obtained from~\eqref{linf} by the substitution
 $C_1=1$,  $C_2=-1$  and  $C_3=0$.
In the case of the sharp cut-off one has
 \begin{align}
\label{linf_s}
  l^{\star}(x)   & = -\ln b_l  -  \beta_l \, \ln x - \frac{1}{b_l} x^{-\beta_l}   -\nonumber \\
                      & - \frac{4}{ b_l \, d \, (d-2)} \, x^{-\beta_l -2 }  +\mathcal{O}(x^{- \beta_l -4}) \; .
 \end{align}
\end{subequations}
In all cases the asymptotics depend on a single parameter $b_l$ which fixes the two initial conditions
 $l^{\star}(x)$ and $ l^{\star \, '}(x)$ at $x=x_{\textrm{max}}$. $b_l$ is determined in such a way that 
 $ l^{\star \, '}(0)=0$. This is the shooting method \textit{ad originem}.

\begin{table}[h]
\caption{\label{tab_b} Results for the coefficient $b_l^*$.}
\begin{tabular}{|l|r|r|r|} 
\hline
                                       &       Litim                                &      Sharp                            &   Smooth                    \\ \hline
$ b_l ^*               $          &    33.3250777220334    &  .091029082436564     &   392.879344670467              \\ \hline
\end{tabular}
\end{table}

The code DOPRI853 detects a stiff problem at large ''x`` so it is impossible to choose  a very large value
of $x_{\textrm{max}}$.  Actually we retained  $x_{\textrm{max}}=15$, $22$ and $20$ for Litim's, the sharp and smooth
cut-off regulators respectively.
However taking into account the full asymptotic expansions\eqref{linf} or \eqref{linf_s}
one recovers exactly, \textit{i.e.} with all the significant figures reported in table~\ref{tab_ab} , the values of $l^{\star}(0)$
obtained by the shooting \textit{ab origine}. The values of coefficient $b_l^*$ at the fixed point obtained
by a dichotomy are given in table~\ref{tab_b}. 
Stiff integrators could be used instead of DOPRI853.

Figure~\ref{l_fix} displays our results for  $l^{\star}(x)$ (smooth cut-off case) obtained by the two shooting methods.
At small ''x''  both curves coincide quasi exactly up to some $\overline{x}$; at $\overline{x}$ two branches of the solution separate one
that comes from the origin,  the other, with scaling properties at infinity which comes from infinity.  This hysteresis phenomenom 
cannot be discarded numerically.

\begin{figure}[t!]
\includegraphics[angle=0,scale=0.35]{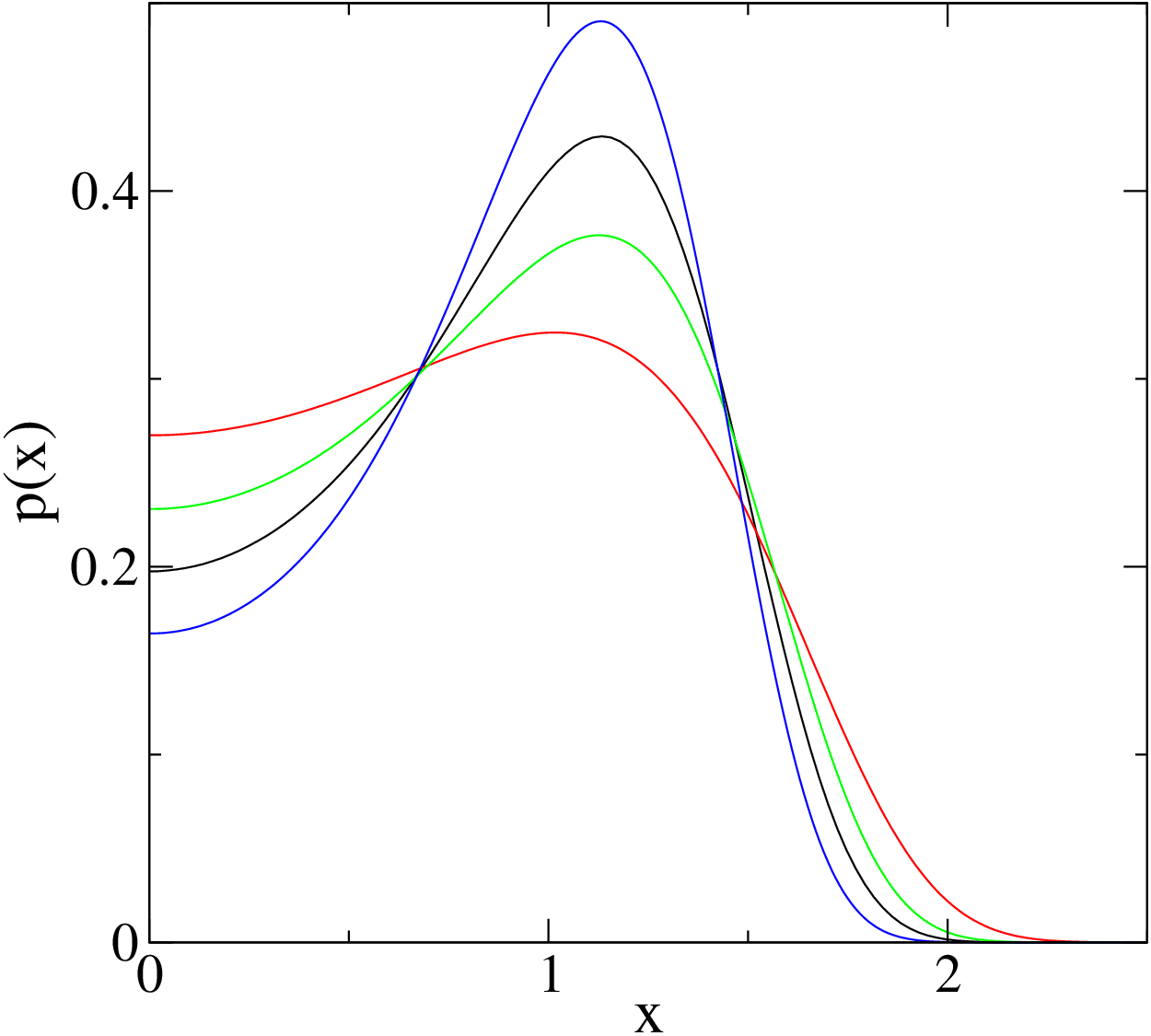}
\caption{
\label{histo} 
Histogram $p(x)$ of the order parameter of the (d=3) Ising model at the critical point.
Black : MC data ~\cite{Tsypin}, red : LPA (Litim), green : LPA (Sharp),
blue : LPA (Smooth)
}
\end{figure}

From  $l^{\star}$  one deduce $u^{\star \, ''} = \omega (l^{\star})$ and thus the fixed point $u^{\star}$ by integration
(up to an additional constant).
In private and informal discussions the opinion circulates that  the
histogram of the order parameter $p(x)$ of the $3d$  Ising model at its critical point
should coincide with $\exp(-u^{\star}(x))$ (up to normalization constants). We are not aware
of any rigorous proof of this assertion but tentatively took it seriously.
This histogram has been obtained by Monte Carlo simulations \cite{Tsypin} and is compared in figure\ref{histo}
with the 
theoretical predictions of the LPA within the three versions considered in this paper. 
Note that all the histograms
has been normalized in such way that the two first even moments are equal to unity, \textit{i.e.}
$\int\, dx \,  p(x)=1$ and $\int\, dx \,  p(x) \, x^2=1$. 
To inject some  quantitative elements in the discussion of
these curves we note that the kurtosis
$K = \int\, dx \,  p(x) \, x^4$ is $K=1.60399$ experimentally, while in the LPA one finds $K=1.86162$, $1.70294$ and
$1.51128$  for Litim's, the sharp and smooth regulators respectively.

The behavior of $p(x)$ at  large deviations ``$x$''  can also be obtained in the framework of the LPA by computing
the asymptotics of the fixed point solution $u^*(x)$. Assuming a power law behavior as $x \to + \infty$,
Eq.~\eqref{fixu} is used to obtain
\begin{subequations}
 \begin{align}
\label{uinf}
  u^{\star}(x)  &=  b_u^* \, x^{\beta_u} + \frac{K_1}{b_u^*} \, \dfrac{(d-2)^2}{2 d (d+2)^2}\,x^{-( \beta_u -2)}  +\nonumber \\
                   & +   \frac{K_2}{b_u^{* \, 2}} \, \dfrac{(d-2)^4}{(2 d)^2 (d+2)^4 (d+4)} \,x^{-2 (\beta_u -2)}
                 +\ldots \; ,
 \end{align}
with $\beta_u= \beta_l +2= 2d/(d-2)$ and $b_u^*$ 
a coefficient which enters at each order in the asymptotics $u^*(x)$ (it's value is such that
$ u^{\star \, '}(0)=0$).
 Of course $b_u^*$ is related to the coefficient $b_l^*$ which governs the asymptotics of
 $l^{\star}(x)$, one finds that   \mbox{$b_u= K_1 (d-2)^2/(b_l^* 2d (d+2))$.} 
Recall finally that $K_1, \, K_2 $ enters the asymptotic behavior~\eqref{asympto2}
of $l=\mathcal{L}(\omega)$ as $\omega  \to + \infty $ and are given in table~\ref{constantes}.

Litim's case can be obtained from~\eqref{uinf} by the substitution
$K_1=1$ and   $K_2=-1$ while, 
in the case of the sharp cut-off,  one has
 \begin{align}
\label{uinf_s}
  u^{\star}(x)   & = b_u^* \, x^{\beta_u} -  \frac{4}{d (d-2)}\, \ln x+ \nonumber \\
                      & + \left( 2 + \ln (b_u^* \beta_u (\beta_u -1)) \right)/d \; ,
 \end{align}
\end{subequations}
where, once again $\beta_u= 2d/(d-2)$ but $b_l^* = b_u^* 2d (d+2)/(d-2)^2$.
\subsubsection{\label{vp} \textit{critical exponents}}
\begin{table*}[t!]
\caption{\label{tab_lambda} Results from the critical exponents of the LPA.}
\begin{tabular}{|l|r|r|r|} 
\hline
                                       &       Litim                             &      Sharp                            &   Smooth                  \\ \hline
$  \lambda_1       $          &    1.539499459806177       &   1.450412451707412       &     1.53706               \\ \hline
$  \nu                   $         &    0.649561773880648       &   0.689459056162135        &   0.650594               \\ \hline
$ \omega_1         $         &    0.655745939193339       &   0.595239852232561        &   0.654104               \\ \hline
$ \omega_2         $         &    3.180006512059168       &   2.838426658241768        &   3.17183                  \\ \hline
$ \omega_3         $         &    5.912230612747701      &   5.184192105359884        &    5.55112                  \\ \hline
$ \omega_4         $         &    8.796092825413904       &   7.596792580450411        &   8.76972                  \\ \hline
$ \omega_5         $         &  11.798087658336857       & 10.057968960649436        & 11.7609                    \\ \hline
$ \omega_6         $         &  14.896053175688298       & 12.556722422589013         &14.8473                 \\ \hline
$\widetilde{\lambda}_2$  &  0.4999999999999999      &  0.499999999999999          &  0.500004             \\ \hline
$\widetilde{\omega}_1$  &   1.8867038380914204       &  1.691338925641807         & 1.88197                \\ \hline
$\widetilde{\omega}_2$  &   4.5243907336707728       &  3.998514715824934        &  4.51219             \\ \hline
$\widetilde{\omega}_3$  &   7.3376506433543136       &  6.382503789820088          & 7.31630                  \\ \hline
$\widetilde{\omega}_4$  & 10.2839007240259583       &  8.821709390384049        & 10.2522                               \\ \hline
$\widetilde{\omega}_5$  & 13.3361699643459432       & 11.302996690411253       &  13.2933                              \\ \hline
\end{tabular}
\end{table*}

We turn now to the eigenvalue equation~\eqref{vp_l}. Again this is a second order ODE the solution of which
is characterized \textit{a priori} by two integration constants. Actually one of these is irrelevant and corresponds to the
arbitrariness of the normalization of an eigenfunction. Since the fixed point solution $l^{\star}(x)$ is an even function of $x$
equation~\eqref{vp_l} is invariant under a parity change and the spectrum separates into even and odd eigenvalues. The 
second integration constant is thus fixed by the choice $g^{'}(0)=0$ (even) or $g(0)=0$ (odd). The shooting ad originem
method seems mandatory and one proceeds as in section~\ref{ad}. 

For the smooth cut-off regulator the asymptotic behavior of the eigenfunctions as $x \to + \infty$ is found to be
\begin{subequations}
 \begin{align}
  g(x)& \sim S \, ( x^{\alpha_l} + 2\frac{C_2}{C_1} _, b_l^* \, x^{\alpha_l - \beta_l}  + \nonumber \\ 
       & 3 b_l^{\star \, 2} \,
        (\frac{C_2^2}{C_1^2} + \frac{C_3}{C_1}) \, x^{\alpha_l -2 \beta_l} + \ldots \; ) ,
 \end{align}
where $S$ is an arbitrary constant and $\alpha_l=(\lambda +2)/(1-d/2)$. Imposing this  form for 
$g(x)$ at some large $x_{\textrm{max}}$ one tunes $\lambda$ to obtain either $g^{'}(0)=0$ or $g(0)=0$ 
The case of the sharp cut-off is special :
 \begin{align}
   g(x) \sim S \, ( x^{\alpha_l} -\frac{x^{\alpha_l - \beta_l}}{b_l^*}+ \ldots \; ) \, ,
 \end{align}
where  $\alpha_l=\lambda/(1-d/2)$.
\end{subequations}

Our results for the even and odd spectra are reported in table~\ref{tab_lambda}. In the even
case there are no trivial eigenvalues, as shown in Sec.~\ref{trivial}, and the first eigenvalue $\lambda_1$
is related to the critical exponent $\nu=1/\lambda_1$ of the correlation length of the Ising model.
The first negative eigenvalue $\lambda_2$ is minus the Ising-like first correction-to-scaling exponent
$\omega_1= -\lambda_2$ and so-on.  In the odd
case, as discussed in Sec.~\ref{trivial} there are in general two positive trivial eigenvalues
 $\widetilde{\lambda}_1$ and  $ \widetilde{\lambda}_2$,
among  which only the second one
$\widetilde{\lambda}_2=d/2 -2 =0.5$ survives in the spectrum. The first non-trivial eigenvalue is negative and 
defines the subcritical exponent $\theta_5=\widetilde{\omega}_1=-\widetilde{\lambda}_3 $ and so on.

The numerical data for $\widetilde{\lambda}_2=0.5$ serves as a stringent test for the precision of the numerical procedure;
while (at least) $15$ significative 
digits are obtained for the sharp and Litim's cut-off no more than $6$ digits can be ascertain in the case of the smooth cut-off.
It originates in the various fitting procedures devised to evaluate the function $\omega(l)$
and its derivatives.  In the case of Litim's and the sharp cut-off our results are in perfect agreement with
those of ref.\cite{BerGiacoI,BerGiacoII}. Overall good agreement between the Litim and smooth cut-off  spectra 
should be stressed.

Since $\eta=0$ in the LPA,  one can compute all critical exponents from the critical exponent of the correlation length $\nu$ by the scaling relations.
One has for $d=3$ : $\alpha = 2 - 3 \nu $, $\beta= \nu/2 $, and $\gamma=2 \nu$. The values are reported in table~\ref{last}.
The comparison with the data of Table~\ref{tab_ind}, obtained by solving the dimensionned PDE flow equation, is deceptive since 
there is no good agreement even by taking into account the error bars. The values obtained for  $\beta$ ( $\gamma$ )
by the numerical experiments of section~\ref{numerI} are systematically larger (smaller) than that obtained in this section. 
The explanation of this discrepancy relies  probably in  systematic errors due to the size of the field and time
steps in the numerical resolution of the PDE.

\begin{table}[h!]
\caption{\label{last} Critical exponents in the LPA}
\begin{tabular}{|l|l|l|l|} 
\hline
                                    &       Litim                                   &      Sharp                                      &   Smooth                  \\ \hline
$   \beta      $               &      0.324780886940324          &     0.3447295280810675             &  0.325297                \\ \hline
$   \gamma $               &      1.299123547761296           &     1.37891811232427                 &  1.301188                \\ \hline
$   \alpha    $               &      0.051314678358056           &   -0.068377168486405                &  0.048218               \\ \hline
\end{tabular}
\end{table}

\section{\label{Conclu} Conclusion}

In this paper we attempted to make an exhaustive study of the properties of the $\varphi^4$
model in the ordered phase in the framework of the NPRG within the LPA approximation. 

We shown that the approach to the convexity is independent of the cut-off,  but that fine details are strongly affected
by the choice of the regulator, notably   the analytical behavior of the inverse magnetic susceptibility
 $\chi^{-1}(M) $ at $M= \pm M_0$. We proved that 
the singularity of the threshold function $\mathcal{L}(\omega)$ about its largest pole (or essential singularity) $\omega_0$
governs the behavior of  $\chi^{-1}(M)$ at $M= \pm M_0$; if
$\mathcal{L}(\omega) \sim (\omega - \omega_0)^{-\nu} \, $  as $\omega \to \omega_0 +$ then  
the inverse compressibility
is discontinuous at $M=\pm M_0$ (as required on physical grounds)  only if the inequality $4 -d -2 \nu \leq 0$ holds.
In particular Litim's and any smooth cut-off yield a discontinuity of  $\chi^{-1}(M)$
in dimension $d=3$, while the sharp cut-off incorrectly predicts a continuous behavior and thus a merging of the spinodal and
binodal curves.

We have confirmed these subtle properties of the solution below $r_c$ by extensive numerical experiments
with the help of a new algorithm which solves the RG  flow equations for the threshold functions rather than for 
the potential or one of its derivatives.  The main advantage of the 
method is to replace the numerical resolution of a highly non linear PDE which exhibits numerical instabilities 
in the ordered phase by that of a quasi-linear parabolic  PDE with good convergence properties.

The ''standard'' version of the LPA retained in this work does not allow to compute the critical parameter  $r_c$  of the model
 which depends strongly upon the choice of cut-off. The main reason is that the choice of the MF approximation as an
initial condition for the RG flow is a too crude approximation. Modified versions of the NPRG ~\cite{ReattoI,ReattoII,Reatto0,Caillol_RG,Dupuis}  remedy to this flaw 
and yield a quite precise estimate  for $r_c$.
However we noticed that, choosing as a new variable $r-r_c$ instead of $r$, the thermodynamics
of the LPA (spontaneous magnetization, magnetic susceptibility) for the $\varphi ^4$  potential
is remarkably independent of the cut-off,
except very close to the critical point.
The latter merely shift $r_c$ to incorrect values.
The numerical solution of the dimensionned RG flow equation does not yield  a very precise estimate of the critical exponents
either, probably because of small numerical errors in the resolution of the PDE.

In order to compute precisely the critical exponents one must 
solve the linearized RG about  the Fisher fixed point. 
It can also be done for the threshold functions instead of the potential or its derivatives without any noticeable
numerical differences but with the   advantage of  getting  rid of some trivial solutions corresponding to 
redundant operators.
The solution of the resulting fixed point equations and associated eigenvalue problems 
can be obtained  with a high numerical precision
with the help of  a non-stiff solver like DOPRI853 \cite{Hairer} for instance. Since the ad originem problem is stiff,
stiff integrators could be of some help however to reduce the integration step and should be tested.

It is not clear whether the LPA scenario for the ordered phase survives for more elaborate approximation schemes, 
this could be the subject of further investigations.

\begin{acknowledgments}
The author would like to thank personally C.~Bervillier for enlightening e-mail
correspondence, notably sec.\ \ref{trivial} owes much to his remarks,  and, collectively,  all the members of the ``groupe de travail NPRG'' 
of the LPTMC (Jussieu, Paris), directed and animated  by G.~Tarjus, 
for many discussions. The anonymous referee of this paper is acknowleged for  many pertinent remarks on the manuscript.
\end{acknowledgments}



\end{document}